\documentstyle[preprint,aps,epsf]{revtex}  
\tightenlines
\begin{document}
\draft


\title {\null\vspace*{-.0cm}\hfill {\small 
hep-ph/9906291} \\ \vskip 0.8cm The Soft Gluon Emission Process in the
Color-Octet Model\\ for Heavy Quarkonium Production}

\author{Cheuk-Yin Wong }

\address{ Physics Division, Oak Ridge National Laboratory, Oak Ridge,
TN 37831}

\maketitle

\begin{abstract}
{ The Color-Octet Model has been used successfully to analyze many
problems in heavy quarkonium production.  We examine some of the
conceptual and practical problems of the soft gluon emission process
in the Color-Octet Model.  We use a potential model to describe the
initial and final states in the soft gluon emission process, as the
emission occurs at a late stage after the production of the heavy
quark pair.  It is found in this model that the soft gluon M1
transition, ${}^1 S_0^{(8)} \rightarrow {}^3 S_1^{(1)} $, dominates
over the E1 transition, ${}^3 P_J^{(8)}\rightarrow {}^3 S_1^{(1)}$,
for $J/\psi$ and $\psi'$ production. Such a dominance may help resolve
the questions of isotropic polarization and color-octet matrix
element universality in the Color-Octet Model.  }

\end{abstract}



\def\bbox#1{\hbox{\boldmath${#1}$}}

\pagebreak

\narrowtext

\section{ Introduction } 

Bodwin, Braaten, and Lepage \cite{Bod95} have recently developed the
Color-Octet Model based on nonrelativistic quantum chromodynamics for
very massive quarks which allows a systematic calculation of inclusive
heavy quarkonium production cross sections.  It is assumed that in
addition to the production of bound states by the color-singlet
mechanism \cite{Cha80}, bound states are produced by the color-octet
mechanism whereby a $Q\bar Q$ pair in a color-octet state is first
formed either by gluon fragmentation or by direct parton reactions,
and the color of the pair is neutralized by emitting a soft gluon of
low energy and momentum.  The cross section for bound state $\psi$
production is represented by
\begin{eqnarray}
\label{eq:octet1}
\sigma( \psi )=\sum_{ab} \sum_{c,J,L,S} F(ab\rightarrow
{}^{2S+1}L_J^c~) 
{\langle0| O({}^{2S+1}L_J^c \rightarrow \psi~)|0\rangle \over M_Q^{d}}
\end{eqnarray} 
where $ab=gg,q\bar q,..$ are the parton combinations leading to the
production, and $c=1,8$ are the color states of the $\psi$ precursor.
The quantity $F(ab\rightarrow {}^{2S+1}L_J^c~)$ is the short-distance
cross section for the production of a $Q\bar Q$ pair with quantum
numbers ${}^{2S+1}L_J^c$ calculated from PQCD, based on the Feynman
diagram for $ab\rightarrow (c \bar c)_{cJLS} +X$.  The matrix element
$\langle0| O({}^{2S+1}L_J^c \rightarrow \psi)|0\rangle$ is the
long-distance probability per unit volume for ${}^{2S+1}L_J^c$ to
produce the final bound state $\psi$ and $d$ is the scaling dimension
of the operator \cite{Bod95}.  For example, for production from the
color-singlet state ${}^{2L+1}L_J^{(1)}$ to the bound state
$\psi_{JLS}$, which has the radial wave function $R_{JLS}(r)$, the
matrix element is
\begin{eqnarray}
\langle0| O_1^{\psi_{JLS}}({}^{2S+1}L_J^{(1)} \rightarrow \psi_{JLS})|0\rangle 
~~\propto~~\left |  \left [ {d ^L R_{JLS}(r) \over dr^L} 
\right ]_{r\rightarrow0} \right  |^2.
\end{eqnarray} 
For production from the color-octet state ${}^{2S'+1} {L'}_{J'}^{(8)}$
to the color-singlet bound state $\psi_{JLS}$, the color-octet matrix
element $\langle {\cal O}_8^{\psi_{JLS}} ({}^{2S'+1} {L'}_{J'})
\rangle$ specifies the probability per unit volume for the $Q\bar Q$
color-octet state ${}^{2S'+1} {L'}_{J'}^{(8)}$ to emit a soft gluon in
the transition to the bound color-singlet state $\psi_{JLS}$.  As the
soft gluon emission takes place on a nonperturbative QCD time scale
and involves nonperturbative QCD, the color-octet matrix elements have
been treated as phenomenological parameters
\cite{Bra95,Tan96,Cho96,Ben96,Bra96,Cac97a,Cac97,Gup96,Pet98}.  Matrix
elements have been extracted to yield good agreement with the CDF data
for high $p_T$ heavy quarkonium production in $\bar p p$ collisions at
$\sqrt{s}=$1.8 TeV \cite{Cho96,Abe97} and with the fixed-target data
for $pN$ collisions at energies up to $\sqrt{s}=40$ GeV \cite{Ben96}.

For the Color-Octet Model to be a valid description, one should be
able to describe the process of soft gluon emission in physical terms.
We know the nature of the final state of observed quarkonium, but what
is the nature of the initial color-octet state which emits a soft
gluon?  What is the nature of the soft gluon?  What is the process of
soft gluon emission?  Which one of the gluon emission processes is
more important in $J/\psi$ and $\psi'$ production?  Besides these
conceptual questions, there are also the following practical questions
which arise in the application of the Color-Octet Model.

In the Color-Octet Model for fixed target energies, the dominant
production comes from the fusion of two gluons forming a $c\bar c$
color-octet pair in ${}^1S_0^{(8)}$ and ${}^3 P_J^{(8)}$ states.  The
soft-gluon radiative transition ${}^1S_0^{(8)} \rightarrow
{}^3S_1^{(1)}$ leads to an isotropic angular distribution of decay
muons in the quarkonium rest frame; the ${}^3 P_J^{(8)} \rightarrow
{}^3S_1^{(1)}$ transition in contrast preferentially populates
$J_z=\pm 1$ substates with a large transverse polarization, which
leads to anisotropic angular distributions in the decay of the
charmonium to muons.  The muon angular distribution can be inferred
from the magnitudes of the matrix elements $\langle {\cal O} _8 ^H
({}^1 S_0) \rangle$ and $\langle {\cal O} _8 ^H ({}^3 P_0)
\rangle/m_c^2$. The $J/\psi$ and $\psi'$ production cross sections at
fixed-target energies give the matrix elements \cite{Ben96}
\begin{eqnarray}
\label{eq:1}
\langle {\cal O} _8 ^{J/\psi} ({}^1 S_0) \rangle + {7 \over m_c^2} 
\langle {\cal O} _8 ^{J/\psi} ({}^3 P_0) \rangle = 3.0 \times 10^{-2}
{\rm~ GeV}^3
\end{eqnarray}
for $J/\psi$ production and 
\begin{eqnarray}
\label{eq:2}
\langle {\cal O} _8 ^{\psi'} ({}^1 S_0) \rangle + {7 \over m_c^2} 
\langle {\cal O} _8 ^{\psi'} ({}^3 P_0) \rangle = 0.5 \times 10^{-2}
{\rm~ GeV}^3
\end{eqnarray}
for $\psi'$ production.  On the other hand, the velocity counting rule
of \cite{Bod95} gives an $\langle {\cal O} _8 ^H ({}^1 S_0) \rangle$
of the same order as $\langle {\cal O} _8 ^H ({}^3 P_0)
\rangle/m_c^2$.  If we set $\langle {\cal O} _8 ^H ({}^1 S_0) \rangle
= \langle {\cal O} _8 ^H ({}^3 P_0) \rangle/m_c^2$, as suggested by
the velocity counting rule of \cite{Bod95}, the expected angular
distribution of muons from $J/\psi$ decay will not be isotropic
\cite{Ben96}.  The experimental data give an isotropic distribution
for $J/\psi$ and $\psi'$ production in $\pi$-W collisions at 252 GeV
\cite{Hei91,Bii87} and 125 GeV \cite{Ake93}.

For the Color-Octet Model to be a valid description, the color-octet
matrix elements $\langle {\cal O} _8 ^{H} ({}^{2S'+1} {L'}_{J'})
\rangle$ should be independent of the processes which produce
${}^{2S'+1} {L'}_{J'}^{(8)}$.  However, there are unresolved questions
concerning this universality of the color-octet matrix elements.  High
$p_T$ CDF measurements at $\sqrt{s}=$1.8 TeV give \cite{Cho96}
\begin{eqnarray}
\label{eq:Cho1}
\langle {\cal O} _8 ^{J/\psi} ({}^1 S_0) \rangle + {3 \over m_c^2} 
\langle {\cal O} _8 ^{J/\psi} ({}^3 P_0) \rangle = 6.6 \times 10^{-2}
{\rm~ GeV}^3
\end{eqnarray}
for $J/\psi$ production and
\begin{eqnarray}
\label{eq:Cho2}
\langle {\cal O} _8 ^{\psi'} ({}^1 S_0) \rangle + {3 \over m_c^2}
\langle {\cal O} _8 ^{\psi'} ({}^3 P_0) \rangle = 1.8 \times 10^{-2}
{\rm~ GeV}^3
\end{eqnarray}
for $\psi'$ production.  On the other hand, $J/\psi$ and $\psi'$
production cross sections at fixed-target energies give the matrix
elements \cite{Ben96} listed in Eqs.\ (\ref{eq:1}) and (\ref{eq:2}).
To check the universality of matrix elements, we need a relation
between $\langle {\cal O} _8 ^{H} ({}^1 S_0) \rangle$ and $\langle
{\cal O} _8 ^{H} ({}^3 P_0) \rangle$ for $H=\{ J/\psi, \psi'\}$.  If
we again use the velocity counting rule of \cite{Bod95} to set
$\langle {\cal O} _8 ^H ({}^1 S_0) \rangle = \langle {\cal O} _8^H
({}^3 P_0) \rangle/m_c^2$, the fixed-target matrix elements are a
factor of 4(7) smaller than the CDF matrix elements for
$J/\psi$($\psi'$), as pointed out by Beneke $et~al.$ \cite{Ben96}.

We would like to formulate the Color-Octet Model in a form which will
allow us to answer these conceptual and practical questions.  We know
that the final observed state can be described nonperturbatively in
terms of a potential model.  Since the emission of the soft gluon
takes place at a long-time scale after the production of the $Q\bar Q$
pair, it is reasonable to describe the initial state which emits a
soft gluon also in terms of a potential model.  The soft-gluon
emission matrix element can then be evaluated with wave functions of
the initial and final states in this potential model and the density
of color-octet states.  From these formulations, we find that when a
color-octet $Q \bar Q$ pair in the ${}^1 S_0^{(8)}$ or ${}^3
P_J^{(8)}$ state emits a very soft gluon to make a transition to the
bound $H={}^3S_1^{(1)}$ state, the probability for the M1 radiative
transition ${}^1 S_0^{(8)} \rightarrow {}^3 S_1^{(1)} $ is much
greater than that for the E1 transition, ${}^3 P_J^{(8)}\rightarrow
{}^3 S_1^{(1)}$, and the velocity counting rule breaks down.  We
suggest that such a dominance of $\langle {\cal O} _8 ^{{}^3S_1} ({}^1
S_0) \rangle$ over $\langle {\cal O} _8 ^{{}^3S_1} ({}^3 P_J)
\rangle/m_c^2$ for soft gluon radiation may help resolve the above
questions of quarkonium polarization and matrix element universality.

This paper is organized as follows.  In Section II, we formulate the
model of quarkonium production in terms of the short-distance Feynman
amplitude and wave functions determined from potential models.  An
explicit expression is obtained to relate the color-octet matrix
element in the Color-Octet Model to quantities which can be evaluated
in the potential model.  In Section III, we examine the potential
between $Q$ and $\bar Q$ interacting at large distances.  When one
allows for the effect of the spontaneous production of light quarks to
break up the string joining $Q$ and $\bar Q$, the linear potential is
modified to become a screened potential.  We use a screened potential
to discuss bound states in Section III and continuum resonance states
in Section IV.  Section IV also outlines how we obtain continuum wave
functions which are needed in the calculation of the cross section.
Section V gives the multipole transition matrix elements.  As an
illustration, a simple M1 to E1 ratio is obtained for the production
of $J/\psi$ using approximate wave functions.  Numerical results with
more realistic potentials are obtained in Section VI to give the
contributions from M1 to E1 transitions to $J/\psi$ and $\psi'$
production.  It is found that the soft gluon M1 radiative transition
${}^1 S_0^{(8)} \rightarrow {}^3 S_1^{(1)} $ dominates over the E1
transition ${}^3 P_J^{(8)}\rightarrow {}^3 S_1^{(1)}$ for $J/\psi$
and $\psi'$ production.  Section VII summarizes and concludes the
present discussions.

\section{ Heavy Quarkonium Production from parton collisions } 

To obtain the cross section for the production of quarkonium bound
states in a hadron-hadron collision, it suffices to focus on the
production cross section in parton collisions, as the former can be
obtained from the latter by folding the parton distribution functions
of the colliding hadrons.  We consider the collision of the parton $a$
with the parton $b$ which form the initial $Q\bar Q$ pair.  The heavy
quarkonium production amplitude can be obtained by projecting out the
state vector of the $Q\bar Q$ pair onto the quarkonium bound state.
We shall show in detail how this is carried out to yield the
quarkonium production cross section.

The production of a heavy quark pair is a fast process.  We can follow
the time evolution of the state vector of the $Q\bar Q$ pair.  The
state vector resulting from the collision of $a$ and $b$ at initial
production time $t_i$ is
\begin{eqnarray}
\label{eq:Phi}
|\Phi_{a b}(t_i) \rangle = |\Phi_{a b}^{Q\bar Q}(t_i) \rangle+ |\Phi_{a
b}^{Q\bar Q g}(t_i)\rangle  + ... 
\end{eqnarray} 
where
\begin{eqnarray}
\label{eq:8}
|\Phi_{a b}^{Q\bar Q}(t_i) \rangle
= \int {d^3 Q  \over (2\pi)^3 {2E_Q} }
 { d^3 {\bar Q} \over (2\pi)^3 {2 E_{\bar Q}}}
(2\pi)^4 \delta^{(4)}(P_i-P_f) {\cal M}(ab \rightarrow Q{\bar Q} ) |Q{\bar
Q} \rangle ,
\end{eqnarray} 
\begin{eqnarray}
\label{eq:9}
 |\Phi_{ab}^{Q\bar Q g}(t_i)\rangle
=  \int {d^3 Q  \over (2\pi)^3 {2E_Q} }
 { d^3 {\bar Q} \over (2\pi)^3 {2 E_{\bar Q}}}
 { d^3g  \over (2\pi)^3 {2E_g}}
(2\pi)^4 \delta^{(4)}(P_i-P_f)
{\cal M}(ab \rightarrow Q{\bar Q} g ) |Q{\bar Q} g
\rangle.
\end{eqnarray} 
Here, $P_i$ and $P_f$ are the initial and final 4-momentum. The matrix
elements $\cal M$ can be obtained in perturbative QCD using Feynman
diagrams and renomalization procedures.

For a sufficiently large value of $t$, the initial state in Eq.\
(\ref{eq:Phi}) will evolve to become
\begin{eqnarray}
\label{eq:Phip}
|\Phi_{a b}(t) \rangle = 
\sum_x 
\int d\Phi_{\psi x}~(2\pi)^4 \delta^{(4)}(P_i-P_f) 
{\cal M}(ab \rightarrow \psi_{JLS} ~x )
|\psi_{JLS}~x \rangle ,
\end{eqnarray} 
where $x$ denotes the number of hard gluons, and $d \Phi_{\psi x}$ is the
corresponding phase space volume element.  For example, for
$ab\rightarrow \psi$, $d\Phi_\psi$ is
\begin{eqnarray}
d\Phi_\psi={d^3 P_\psi   \over (2\pi)^3 {2E_\psi} },
\end{eqnarray} 
and for $ab\rightarrow \psi~g$ with the emission of a hard gluon,
$d\Phi_{\psi g}$ is
\begin{eqnarray}
d\Phi_{\psi g}={d^3 P_\psi   \over (2\pi)^3 {2E_\psi} }
 { d^3g  \over (2\pi)^3 {2E_g}}.
\end{eqnarray} 
The cross section for the production of the quarkonium state
$\psi_{JLS}$ by the collision of partons $a$ and $b$ is then
\begin{eqnarray}
\label{eq:xsec}
d\sigma&(ab\rightarrow \psi_{JLS}~x ) = {1 \over 4 I} (2 \pi)^4
 \delta^{(4)}(P_i-P_f) \sum\limits_{x} |{\cal M }(ab\rightarrow \psi_{JLS}~x
 )|^2 d\Phi_{\psi x}
\end{eqnarray}
where $I= \sqrt{(p_a\cdot p_b)^2 - m_a^2 m_b^2}$.  It is necessary to
obtain ${\cal M }(ab\rightarrow \psi_{JLS}~x )$ from ${\cal M
}(ab\rightarrow Q\bar Q~x )$ in order to calculate the production
cross section.  We shall show how this relation can be obtained in the
present model description.

One can first separate out the center-of-mass motion and the relative
motion of the $Q \bar Q$ pair in terms of their total momentum $P$ and
their relative momentum $q$. The heavy quark momentum is then
$Q=P/2+q$, the antiquark momentum is ${\bar Q}=P/2-q$, and the state
of the center-of-mass motion can be represented by the plane wave
state $|Q\bar Q \rangle$ which is equivalent to $|q\, P \rangle$.
They are related by the ratio of their normalization constants.

We look first at the situation when $Q$ and $\bar Q$ are not
interacting to introduce their interaction through a mutual potential
$V$ later on.  Part of the phase space elements, $d^4Qd^4\bar Q$, in
Eqs.\ (\ref{eq:8}) and (\ref{eq:9}) can be transformed into
\begin{eqnarray}
\label{eq:14}
 {d^4 Q \over (2\pi)^3} \delta(Q^2-m_Q^2) { d^4 {\bar Q} \over
 (2\pi)^3} \delta({\bar Q}^2-m_{\bar Q}^2) = {d^4 P \over (2\pi)^3}
 \delta(P^2-m_P^2) { d^4 { q} \over (2\pi)^3} \delta \left ( {P\cdot q
 \over 2} \right )
\end{eqnarray} 
where $m_P$ is the invariant mass of the $Q \bar Q$ system.  To study
the dynamics of the relative momentum, it is best to refer to the
center-of-mass system where $P=(m_P,0)$, $q=(0,\bbox{q})$, and the
relative momentum $\bbox{q}$ satisfies the following equation
\begin{eqnarray}
\label{eq:NI}
\epsilon_\omega^2-\bbox{q}^2-m_\omega^2=0,
\end{eqnarray} 
where 
\begin{eqnarray}
\epsilon_\omega={m_P^2 - 2 m_Q^2 \over 2 m_P},
\end{eqnarray} 
and $m_\omega$ is the reduced mass,
\begin{eqnarray}
m_\omega=m_Q^2/{m_P}.
\end{eqnarray} 
Next, we consider the system with a mutual time-like vector interaction
$V$.  The equation of motion can be obtained from Eq. (\ref{eq:NI}) by
the canonical method of replacing $\epsilon_w$ with $\epsilon_w -
V(r)$.  The Klein-Gordon equation for the two-particle system under a
mutual time-like vector interaction $V(r)$ is
\begin{eqnarray}
\label{eq:KG}
\biggl \{ [\epsilon_\omega - V (r) ]^2 - \bbox{q}^2 - m_w^2 \biggr \} 
\Psi(\bbox{r}) = 0.
\end{eqnarray}
Similar two-body equations of constraint dynamics for more complicated
cases with spin and very general types of interactions can be found in
\cite{Cra96}.  Because of the large mass of the heavy quark, it is
convenient to use nonrelativistic $Q \bar Q$ wave functions and state
vectors.  The Klein-Gordon equation becomes the Schr\" odinger
equation,
\begin{eqnarray}
\label{eq:KG1}
\left  \{  {\bbox {q} ^2 \over 2 m_\omega} + V(r) - \epsilon \right \}
\Psi(\bbox{r}) = 0,
\end{eqnarray} 
where $\epsilon=\epsilon_\omega-m_\omega$ is the non-relativistic
measure of energy, and $m_\omega\sim M_Q/2$.  The eigenvalue
$\epsilon$ of the bound state is related to the masses by
$m_P=\epsilon+2m_Q$.  The bound state wave function can be written in
the form
\begin{eqnarray}
\Psi_{JLS} (\bbox{r})= R_{JLS}(r) {\cal Y}_{JLS}(\hat {\bbox{r}}).
\end{eqnarray}
The wave function in momentum space is then
\begin{eqnarray}
{\psi}_{JLS} (\bbox{q})= 
\int d\bbox{r} e^{-i \bbox{q}\cdot \bbox{r}} \Psi_{JLS} (\bbox{r}).
\end{eqnarray}
Heavy-quarkonium bound states have been
obtained previously with many different forms of the potentials
\cite{Eic95}.

In nonrelativistic kinematics, $P=(P_0,\bbox{P})$, and
$q=(q_0,\bbox{q}) \approx(0,\bbox{q}) $.  One can perform a
decomposition of the relative wave function in terms of color and
angular momentum states.  The amplitudes in Eq.\ (\ref{eq:Phi}) can be
decomposed as
\begin{eqnarray}
\label{eq:trans}
 {\cal M}(ab \rightarrow Q{\bar Q} x ) |Q{\bar
Q} \rangle 
=\sum_{c J L S }
{ \cal M}^c_{JLS}( \bbox{q} ~x)  |\bbox{q} ~P \rangle,
\end{eqnarray}
where ${ \cal M}^c_{JLS}( \bbox{q} ~x)$ is
\begin{eqnarray}
{ \cal M}^c_{JLS}( \bbox{q}~x)= {\cal A}^c_{JLS}(|\bbox{q}|~x){ \cal Y}^c_{JLS}(\hat{\bbox{q}})
\end{eqnarray}
with ${ \cal Y}^c_{JLS}(\hat{\bbox{q}})$ the angular momentum and
color wave function.  For simplicity of notation, the azimuthal
angular momentum component $M$ and the color component of the color
multiplet in ${ \cal Y}^c_{JLS}$ will not be written out explicitly.
The bound state $\psi_{JLS}$ can be described as
\begin{eqnarray}
\label{eq:JLSP}
|\psi_{JLS}~P \rangle= \int {d\bbox{q} \over (2\pi)^3}
{ \psi}_{JLS}({ \bbox{q}})
|\bbox{q} ~P \rangle.
\end{eqnarray}
In the lowest order of approximation without soft gluon emission, the
probability amplitude for the direct production of $\psi_{JLS}$ is
obtained by projecting $\psi_{JLS}$ onto $\Phi_{ab}(t_i) $.  We
use the normalization for the center-of-mass motion
\begin{eqnarray}
\langle P'| P \rangle= (2\pi)^3 2E_P \delta(\bbox{P}'-\bbox{P})
\end{eqnarray}
and the usual convention for nonrelativistic relative motion
\begin{eqnarray}
\langle \bbox{q}'|\bbox{ q} \rangle= (2\pi)^3 \delta(\bbox{q}'-\bbox{q}).
\end{eqnarray}
To carry out the projection of $\Phi_{ab}(t_i) $ onto $\psi_{JLS}$, we
use Eq.\ (\ref{eq:trans}) to write Eqs.\ (\ref{eq:8}) and (\ref{eq:9})
as
\begin{eqnarray}
|\Phi_{a b}^{Q\bar Q}(t_i) \rangle
= \int {d^3 Q  \over (2\pi)^3 {2E_Q} }
 { d^3 {\bar Q} \over (2\pi)^3 {2 E_{\bar Q}}}
(2\pi)^4 \delta^{(4)}(P_i-P_f) 
\sum_{c J L S } { \cal M}^c_{JLS}( \bbox{q} ~x)  |\bbox{q} ~P \rangle,
\end{eqnarray} 
\begin{eqnarray}
 |\Phi_{ab}^{Q\bar Q x}(t_i)\rangle
=  \int {d^3 Q  \over (2\pi)^3 {2E_Q} }
 { d^3 {\bar Q} \over (2\pi)^3 {2 E_{\bar Q}}}
 { d^3g  \over (2\pi)^3 {2E_g}}
(2\pi)^4 \delta^{(4)}(P_i-P_f)
\sum_{c J L S }
{ \cal M}^c_{JLS}( \bbox{q} ~x)  |\bbox{q} ~P \rangle.
\end{eqnarray} 
The scalar product of $|\psi_{JLS}~P x \rangle$ with the above state
vector, $\langle \psi_{JLS}~P x |\Phi_{ab}^{Q\bar Q x}(t_i)\rangle$,
can be evaluated by transforming to the variables $P$ and $q$ with
Eq.\ (\ref{eq:14}).  Using Eq.\ (\ref{eq:JLSP}), we obtain
\begin{eqnarray}
\langle \psi_{JLS}~P x |\Phi_{ab}^{Q\bar Q x}(t_i)\rangle
=(2\pi)^4 \delta^{(4)}(P_{ab}-P_\psi)
{ 2 \over m_{JLS}}
\int  { d^3\bbox{q}  \over (2\pi)^3 }
 \psi_{JLS}^*({ \bbox{q}}){\cal M}^{(1)}_{JLS}( \bbox{q}~x). 
\end{eqnarray} 
Therefore, from the above result and the definition of ${\cal M}(ab
\rightarrow \psi_{JLS} ~x )$ as given by Eq.\ (\ref{eq:Phip}), we
get  
\begin{eqnarray}
\label{eq:proj}
{\cal M }(ab\rightarrow \psi_{JLS}~x )= 
{2 \over m_\psi} 
\langle \psi_{JLS}(\bbox{q})|{\cal M}^{(1)}_{JLS}( \bbox{q}~x) \rangle,
\end{eqnarray}
where 
\begin{eqnarray}
\label{eq:proj1}
\langle { \psi}_{JLS}({\bbox{ q}})|{
\cal M}^{(1)}_{JLS}( \bbox{q}~x) \rangle
=\int  { d^3\bbox{q}  \over (2\pi)^3 }
{ \psi}_{JLS}^*({ \bbox{q}}){\cal M}^{(1)}_{JLS}( \bbox{q}~x),
\end{eqnarray}
which involves only color-singlet components of $|\Phi_{a b}(t_i)
\rangle$.  Thus, if the bound state wave function is known, the above
overlap will give the matrix element and the color-singlet
contribution to the quarkonium production cross section in Eq.\
(\ref{eq:xsec}). 

In many approximate calculations one expands ${\cal A}^{(1)}_{JLS}(
|\bbox{q}|~x)$ in powers of the velocity $v=|\bbox{ q}|/M_Q$ and
retains only the leading term, 
\begin{eqnarray}
{\cal A}^{(1)}_{JLS}(|{ \bbox{q}} |~x) \sim 
{\overline {\cal A}}^{(1)}_{JLS x}\times |{ \bbox{q} \over M_Q} | ^L,
\end{eqnarray} 
where 
\begin{eqnarray}
{\overline {\cal A}}^{(1)}_{JLS x}={M_Q^L  \over  L!}\left [ {d^L\over d
|\bbox{ q}|^L} {\cal A}^{(1)}_{JLS}(|\bbox{q}|~x)\right
]_{|\bbox{q}|\rightarrow 0}.
\end{eqnarray} 
Then the above projection in Eqs.\ (\ref{eq:proj}) and
(\ref{eq:proj1}) gives the $L$-th derivative of the wave function at
the origin \cite{Cah87,Pes95,Cra91}:
\begin{eqnarray}
\label{eq:der}
\langle {\psi}_{JLS}({\bbox{ q}})|{\cal M}^{(1)}_{JLS}( \bbox{q}~x) \rangle
/M_Q^{2L}
&=&{ {\overline {\cal A}}^{(1)}_{JLSx} } 
\langle  {\psi}_{JLS}({ \bbox{q}})~|~~ |{ \bbox{q} \over M_Q} |^L 
{ \cal Y}^c_{JLS}\rangle \nonumber \\
&=&  { {\overline {\cal A}}^{(1)}_{JLSx} }  
(-i)^L {(2L+1)!!\over M_Q^L 4 \pi \, L!} 
 \left [ {d^L R_{JLS}(r) \over dr^L} \right ]_{r\rightarrow 0}.
\end{eqnarray} 
The cross section can then be separated into a short-distance part
involving ${\overline A}^{(1)}_{JLS x}$ and the long-distance
nonperturbative part involving the $L$-th derivative of the bound
state wave function at the origin.  We can write the above results in
the notation of the Color-Octet Model of Eq. (\ref{eq:octet1}):
\begin{eqnarray}
\sigma( \psi )=\sum_{ab} \sum_{c,J,L,S} F(ab\rightarrow
{}^{2S+1}L_J^c~) 
{\langle0| O({}^{2S+1}L_J^c \rightarrow \psi~)|0\rangle \over M_Q^{2L}},
\end{eqnarray} 
where
\begin{eqnarray}
\label{eq:octet}
F(ab\rightarrow {}^{2S+1}L_J^{(1)}~)=
\int  {1 \over 4 I} (2 \pi)^4
 \delta^{(4)}(P_i-P_f)
\left ( {2 \over m_\psi} \right ) ^2 
 \sum\limits_{x} 
{L!  |{\overline {\cal A} }_{JLSx}^{(1)} |^2 \over 4 \pi 2 N_c (2L+1) !!}
  d\Phi_{\psi x},
\end{eqnarray}
and following Bodwin $et~al.$\cite{Bod95}
\begin{eqnarray} 
  { \langle 0| O ({}^{2S+1}L_J^{(1)} \rightarrow \psi~)|0\rangle 
   \over M_Q^{2L} }
=  {2 N_c \over 4 \pi} {(2L+1)!!\over  L!}
\left [ {d^L R_{JLS}(r) \over M_Q^L  dr^L} \right ]_{r\rightarrow 0}^2.
\end{eqnarray}

In addition to these color-singlet contributions, the color-octet
model further postulates a much slower process with the emission of a
soft gluon $g_s$, different from the emission of the hard gluon $g$ in
perturbative QCD already included in the second amplitude $|\Phi_{a
b}^{Q\bar Q g}\rangle$ of Eq.\ ({\ref{eq:Phi}).  One can describe the
emission of a soft gluon by studying further evolution of the state
$|\Phi_{a b} (t_i) \rangle$ under the influence of the color field in
the form
\begin{eqnarray}
\label{eq:ev}
|\Phi_{a b}(t) \rangle~ = ~U(t,t_i)|\Phi_{a b}(t_i) \rangle 
\sim \left (1 - ig \int_{t_i}^t~\int {\hat j}\cdot {\hat  A} ~d^3x ~dt
\right )~ |\Phi_{ab}(t_i) \rangle,
\end{eqnarray}
where we shall focus attention on soft gluons only for the gluon field
$A$.

The probability amplitude ${\cal M} (ab\rightarrow \psi_{JLS}~ g_s)$
for the production of the bound state $\psi(JLS)$ accompanied by a
soft gluon $g_s$ is
\begin{eqnarray}
\label{eq:soft}
(2\pi)^4 \delta(P_{ab}-P_\psi-p_{g})
{\cal M} (ab\rightarrow \psi_{JLS}~ g_s)
&=&\langle [\psi_{JLS};P_\psi]~g_s| \Phi_{ab}(t) \rangle \nonumber\\
&=& \langle [\psi_{JLS};Pq]~g_s|~(- ig) \int ~{\hat j}\cdot {\hat  A} 
~d^3x~dt~|
\Phi_{ab}(t_i)\rangle 
\end{eqnarray}
where we have used the notation $g_s$ with the subscript $s$ to denote
a soft gluon emitted by a color-octet state.
Using Eqs.\ (\ref{eq:8}) and (\ref{eq:14}), the righthand side of
the above equation becomes
\begin{eqnarray}
\label{eq:oct0}
& &{\rm RHS~of~(\ref{eq:soft})}=  \sum_{J'L'S'x} \int {d^4 P \over
(2\pi)^3} \delta(P^2-m_P^2) {d^4 q\over (2\pi)^3} \delta\left 
({P\cdot q \over 2} \right )
(2\pi)^4\delta^{(4)}(P_{ab}-P) 
\nonumber\\ 
& &\times \langle (\psi_{JLS}\, P_\psi)
g_s |(- ig) \int_{t_i}^t \int {\hat j} \cdot {\hat A} ~d^3x ~dt
{\cal M}_{J'L'S'}^{(8)}(q~x)|q \, P \rangle.
\end{eqnarray}
   
One envisages that in order to emit a soft gluon, the state
$\Phi_{ab}$ must have evolved for a long time and has characteristics
of nonperturbative states.  It is therefore reasonable to postulate
that the soft gluon is emitted by an intermediate nonperturbative
state $\psi_{J'L'S'}^{(8)}$ best described nonperturbatively in terms
of $Q$ and $\bar Q$ interacting in a potential.  It is appropriate to
expand the intermediate states in terms of $\psi_{J'L'S'}^{(8)}$ with
an invariant mass $m_{J'L'S'}$ using
\begin{eqnarray}
\sum_{J'L'S'x}\int dm_{J'L'S'} {dn\over dm_{J'L'S'}} 
|\psi_{J'L'S'}^{(8)}  \rangle 
\langle\psi_{J'L'S'}^{(8)}| = 1.
\end{eqnarray}
Here $dn/dm_{J'L'S'}$ is the density of these color-octet
$\psi_{J'L'S'}^{(8)}$ states with a specific polarization and
color-octet component:
\begin{eqnarray}
{dn\over dm_{J'L'S'}}={m_Q^{3/2} \over (2\pi)^2} 
\sqrt{m_{J'L'S'}-2m_Q- { L'(L'+1) \over m_QR^2} },
\end{eqnarray}
where $R$ is the average radius of the $Q$-$\bar Q$ separation of the
color-octet state which emits the soft gluon. It can be taken
approximately as the $Q$-$\bar Q$ separation of the final
color-singlet quarkonium state.  For the emission of a soft gluon
through the intermediate color-octet state $\psi_{J'L'S'}^{(8)}$, the
righthand side of Eq.\ (\ref{eq:soft}) is
\begin{eqnarray}
\label{eq:octmat}
& &{\rm RHS~of~(\ref{eq:soft})}= \sum_{J'L'S'x} \int {d^4 P \over
(2\pi)^3}\delta(P^2-m_{J'L'S'}^2) {d^4 q\over (2\pi)^3} \delta\left
({P\cdot q \over 2} \right ) (2\pi)^4\delta(P_{ab}-P) \nonumber\\ &
&\times \langle (\psi_{JLS}\, P_\psi) g_s |(- ig) \int_{t_i}^t \int
{\hat j}\cdot {\hat A} ~d^3x ~dt |\psi_{J'L'S'}^{(8)} \, P \rangle
dm_{J'L'S'} {dn\over dm_{J'L'S'}} \langle\psi_{J'L'S'}^{(8)}|{\cal
M}_{J'L'S'}^{(8)}(q~x)|q\rangle.
\end{eqnarray}
The integral over $q$ can be carried out to give
\begin{eqnarray}
\int {d^4 q\over (2\pi)^3} \delta \left ({P\cdot q \over 2} \right ) 
\langle\psi_{J'L'S'}^{(8)}|{\cal M}_{J'L'S'}^{(8)}(q~x)|q\rangle
&=&{2 \over m_{J'L'S'}}\int  {d^3 q\over (2\pi)^3}
\psi_{J'L'S'}^{(8)*}(q) {\cal M}_{J'L'S'}^{(8)}(q~x) 
\nonumber \\
&\equiv&{2 \over m_{J'L'S'}} 
\langle\psi_{J'L'S'}^{(8)}({\bbox{q}}) 
|{\cal M}_{J'L'S'x}^{(8)} ({\bbox{q}}x) \rangle.
\end{eqnarray}
Thus, the amplitude involves a projection of the color-octet Feynman
amplitude onto a color-octet nonperturbative amplitude,
$\langle\psi_{J'L'S'}^{(8)}(\bbox{q})|{\cal
M}_{J'L'S'}^{(8)}(\bbox{q}\, x)\rangle$, similar to the projection of
the color-singlet component onto a color-singlet amplitude as given in
Eq.\ (\ref{eq:proj}).  One can thus similarly relate this projection
to the $L$-th radial derivative of the spatial wave function, as was
shown in Eq.\ (\ref{eq:der}). One obtains
\begin{eqnarray}
\label{eq:der1}
{\langle {\psi}_{J'L'S'}^{(8)}({ q})|{\cal M}^{(8)}_{J'L'S'}( q~x)
\rangle 
\over M_Q^{L'}}
= { {\overline {\cal A}}^{(8)}_{J'L'S'x} }
(-i)^{L'} {(2L'+1)!!\over M_Q^{L'} 4 \pi \,  {L'}!}\left [ {d^{L'} R_{J'L'S'}(r) \over
dr^{L'}} \right ]_{r\rightarrow 0}
\end{eqnarray}
where
\begin{eqnarray}
{\overline {\cal A}}^{(8)}_{J'L'S' x}={M_Q^L \over  {L'}!}  \left [ {d^{L'}\over d
|\bbox{ q}|^{L'}} {\cal A}^{(8)}_{J'L'S'}(|\bbox{q}|~x) \right
]_{|\bbox{q}|\rightarrow 0}.
\end{eqnarray}

To evaluate the long-distance soft gluon radiation matrix element, we
have
\begin{eqnarray}
\hat A^\mu = \sum_k \sqrt{4 \pi \over 2 \omega_g}~~[{\hat c_k}
\epsilon^\mu  \,
e ^{i ({\bbox {k}} \cdot {\bbox {r}} -\omega_g t)}
+ {\hat c}_k^\dagger 
\epsilon^{\mu *} \,
e ^{-i ({\bbox {k}} \cdot {\bbox {r}} -\omega_g t)}
]
\end{eqnarray}
where $\omega_g$ is the soft gluon energy, $\epsilon^\mu$ is the gluon
polarization vector, $c_k$ and $c_k^\dagger$ are the creation and
annihilation operator for the soft gluon.  Therefore, we have
\begin{eqnarray}
& & \langle (\psi_{JLS}\, P_\psi)~g_s(p_g)|~(- ig) \int_{t_i}^t ~{\hat
j}\cdot {\hat A} ~d^3r~dt~| \psi_{J'L'S'}^{(8)}(q)\, P \rangle \nonumber \\ 
&=& V_{J'L'S'\rightarrow JLS} 
(2\pi)^3 \delta(\bbox{P}-\bbox{P}_\psi-\bbox{P}_g)
\int_{t_i}^{t} e^{-i(E_i-E_f-\omega_g)t}
dt \nonumber  
\end{eqnarray}
where
\begin{eqnarray}
V_{J'L'S'\rightarrow JLS}=-ig \sqrt{4 \pi \over 2 \omega_g} \epsilon_\mu^* 
~\int j_{J'L'S'\rightarrow JLS}^\mu 
(\bbox{r}) e^{-i {\bbox{k}}\cdot {\bbox{r}}}
d \bbox{r},
\end{eqnarray}
the time-like component of $j_{fi}^0(\bbox{r})=\rho_{fi}(\bbox{r})$ is
\begin{eqnarray}
\rho_{fi}(\bbox{r})=
\int d\bbox{r}_1 \, d\bbox{r}_2 \psi_f^*(\bbox{r}_1,\bbox{r}_2)
\sum_{n=1}^2 \delta(\bbox{r}-\bbox{r}_n) 
\psi_i(\bbox{r}_1,\bbox{r}_2)
\end{eqnarray}
where $\bbox{r}_1,\bbox{r}_2$ refer to the coordinates of the heavy
quark and antiquark, and $\psi_i$ and $\psi_f$ are the initial and
final wave functions. 
The space-like component is
\begin{eqnarray}
\bbox{j}_{fi}(\bbox{r})= - {ig \over 2 M_Q }(\psi_f^*(\bbox{r})\nabla
\psi_i (\bbox{r})
-\psi_i^*(\bbox{r})\nabla \psi_f(\bbox{r}) ) 
+ {g \mu_Q \over 2 M_Q} \nabla \times (\psi_f^*(\bbox{r})
{\bbox{\sigma}} \psi_i(\bbox{r})) ,
\end{eqnarray}
where $\mu_Q\sim 2$ is the color-magnetic moment of the quark in units
of the color Bohr magneton.
Therefore, we have 
\begin{eqnarray}
 {\cal M}( {ab\rightarrow \psi_{J'L'S'}^{(8)} x \rightarrow
\psi_{JLS}~ g_s \, x }) 
= \langle \psi_{J'L'S'}^{(8)}(\bbox{q}\,x) | {\cal M}_{J'L'S'}^{(8)}
(\bbox{q})\rangle
 V_{J'L'S'\to JLS} 
 {\sqrt{ 2E_{J'L'S'} }\over  m_{J'L'S'}^2} 2\pi{ dn
\over dm_{J'L'S'}}
\end{eqnarray}
where $E_{J'L'S'}=\sqrt{m_{J'L'S'}^2+(\bbox{a}+\bbox{b})^2}$.  From Eq.\
(\ref{eq:xsec})
\begin{eqnarray}
\label{eq:oct}
d\sigma(ab\rightarrow \psi_{JLS}~ g_s \,x) = {1 \over 4I}
\sum_{xJ'L'S'}
| {\cal M}(
{ab\rightarrow 
\psi_{J'L'S'}^{(8)} x \rightarrow
\psi_{JLS}~ g_s \, x}) |^2 (2\pi)^4 \delta^{(4)}( P_i-P_f) d\Phi_{\psi
g_s x}
\end{eqnarray}
where the phase space volume element $d\Phi_{\psi g_s x}$ includes
both the soft and the hard gluons.  
For the case when only a soft gluon is
emitted, $d\Phi_{\psi g_s}$ is given by
\begin{eqnarray}
d\Phi_{\psi g_s}={d^3 P_\psi   \over (2\pi)^3 {2E_\psi} }
 { d^3g_s  \over (2\pi)^3 {2E_{g_s}}}.
\end{eqnarray} 
We have then 
\begin{eqnarray}
\label{eq:xsecsf}
d\sigma(a b\rightarrow \psi_{JLS} ~g_s ) 
&=& {1 \over 4I}
\sum_{xJ'L'S'}
\left ( {2\over m_{J'L'S'} }\right )^2
|\langle\psi_{J'L'S'}^{(8)}(\bbox{q})|{\cal
M}_{J'L'S'}^{(8)}(\bbox{q}~x)\rangle|^2 \nonumber \\
& & \times |V_{J'L'S'\rightarrow JLS}|^2
\, {1\over 4 m_{J'L'S'}^2 } \left ( 2\pi { dn
\over dm_{J'L'S'}}\right )^2
{|{\bbox{p}}_{\psi}| \over 8\pi^2} d\Omega_\psi.
\end{eqnarray}
where $|{\bbox{p}}_{\psi}|$ is the magnitude of the momentum of
$\psi_{JLS}$ in the rest frame of $\psi_{J'L'S'}^{(8)}$.  One can
expand $\langle\psi_{J'L'S'}^{(8)}(\bbox{q})|{\cal
M}_{J'L'S'}^{(8)}(\bbox{q}~x)\rangle$ out according to Eq.\
(\ref{eq:der1}) and obtain
\begin{eqnarray}
\label{eq:xsecfn}
d\sigma(a b\rightarrow \psi_{JLS} ~g_s ) 
&=& {1 \over 4I}
\sum_{xJ'L'S'}
\left ( {2\over m_{J'L'S'} }  \right )^2
\left | {\overline {\cal A}}^{(8)}_{J'L'S'x}  \right |^2
     {(2L'+1)!!\over 4 \pi \,  {L'}!} 
\left | \left [ {d^{L'} R_{J'L'S'}(r) \over
 M_Q^{L'} dr^{L'}} \right ]_{r\rightarrow 0} \right |^2  \nonumber \\
& & \times |V_{J'L'S'\rightarrow JLS}|^2
\, {1\over 8 m_{J'L'S'}^2 } \left ({ dn
\over dm_{J'L'S'}}\right )^2
{|{\bbox{p}}_{\psi}|} d\Omega_\psi.
\end{eqnarray}
We can then compare the above results with the expressions in the
Color-Octet Model, which we write in the form
\begin{eqnarray}
\label{eq:xseccom}
d\sigma(&a&b\rightarrow \psi_{JLS} ~g_s ) \nonumber\\
&=& {1 \over 4I}
\sum_{J'L'S'}
f(ab\rightarrow J'L'S') (2\pi)^4 \delta^{(4)}(p_a+p_b-p_\psi)
{d^3 p_\psi \over (2\pi)^3 2E_\psi}
{ \langle O_8^{JLS}(J'L'S')\rangle
 \over M_Q^{2L'}} ,
\end{eqnarray}
where
\begin{eqnarray}
f(ab\rightarrow J'L'S') 
={1 \over 4I}
\sum_{J'L'S'}
\left ( {2\over m_{J'L'S'} }\right )^2
 \left | \sqrt{ {(2L'+1)!! \over 4 \pi C_{J'L'S'} L'! }
\overline {{\cal  A}}_{J'L'S'}^{(8)}}  \right |^2,
\end{eqnarray}
where $C_{J'L'S'}=($number of polarization$)\times (N_c^2-1)$ for the
color-octet state $\{J'L'S'\}$.  We observe the following differences.
The Color-Octet Model result of Eq.\ (\ref{eq:xseccom}) assumes that
the soft photon does not carry energy, hence the cross section is a
delta function of the total parton invariant mass, centered at the
invariant mass of $\psi_{JLS}$. The soft gluon makes no contribution
to the phase space volume element and the matrix element is
parameterized as unknown constants.  On the other hand, when one takes
into account the details of the emission process to obtain the results
in Eq.\ (\ref{eq:xsecfn}), the cross section is a continuous function
of the total parton invariant mass $m_{J'L'S'}$. The production of
$\psi_{JLS}$ occurs within the invariant mass interval from the
minimum $m_{J'L'S'}$(min), at which the color-octet state begins to
exist, to the maximum value $m_{J'L'S'}$(max)=$m_{JLS}+\omega_g({\rm
cutoff})$, which corresponds to the maximum gluon energy of the soft
gluon process.  One envisages that gluons with energies $\omega_g$
higher than $\omega_g$(cutoff) will be emitted at a shorter time scale
and will be described by perturbative QCD and not by the soft gluon
emission process through an intermediary non-perturbative state.  As
the time to travel the length of $R/2$ of a $J/\psi$ corresponds to a
energy scale of $2\hbar/R$, this $\omega_g$(cutoff) energy should be
smaller for a final quarkonium state with a greater radius (such as
$\psi'$).  Furthermore, another difference is that the cross section
of Eqs.\ (\ref{eq:oct}) and (\ref{eq:xsecfn}) in the present
description is given in terms of transition matrix elements of
nonperturbative wave functions, and the phase-space volume element
includes the effect of the additional soft gluon.

In an approximate comparison by integrating Eqs.\
(\ref{eq:xsecfn}) and (\ref{eq:xseccom}), we get approximate equivalence
\begin{eqnarray}
\label{eq:come}
{ \langle O^{(8)}(J'L'S' \rightarrow JLS)\rangle \over M_Q^{2L'}}
&\sim&
E_{JLS}
 \int_{{m_{J'L'S'}}({\rm min})}^{{m_{J'L'S'}}({\rm max})} dm_{J'L'S'}
\left | \sqrt{ C_{J'L'S'} (2L'+1)!! \over 4 \pi L'! } 
\left [ {d^{L'} R_{J'L'S'}(r) \over M_Q^{L'}
dr^{L'}} \right ]_{r\rightarrow 0} \right |^2
\nonumber \\
& &\times
|V_{J'L'S'\rightarrow JLS}|^2
\, { 1\over 8\pi  m_{J'L'S'}^2 } \left ({ dn
\over dm_{J'L'S'}}\right )^2
{|{\bbox{p}}_{\psi}|} d\Omega_\psi.
\end{eqnarray}

We shall consider the soft gluon to have large wave lengths which are
greater than the dimension of the radiating system.  In this long wave
length approximation the matrix element $V_{fi}$ has been worked out
in detail (See Eqs.\ (46.6) and (46.7) of Ref. \cite{Ber82}).  The
results for gluon radiation of multipolarity $LM$ are
\begin{eqnarray}
|V_{fi}|^2=4 {(L+1) \over L} {\omega_g^{2L+1} \over [(2L+1)!!]^2}
|M_{LM}^{(e,m)}|^2.
\end{eqnarray}
For electric transition of
multipolarity $LM$, the multipole transition moment is
\begin{eqnarray}
M_{LM}^{(e)}= g
\int d\bbox{r} \rho_{fi}(\bbox{r})  r^L Y_{LM}
({\hat {\bbox{r}}}).
\end{eqnarray}
 For magnetic transition of multipolarity $LM$,
the multipole transition moment is
\begin{eqnarray}
M_{LM}^{(m)}=
{1 \over L+1} \int d\bbox{r}
~\bbox{r} \times \bbox{j}_{fi}(\bbox{r}) \cdot \nabla \left ( r^L Y_{LM}
({\hat {\bbox{r}}})\right ).
\end{eqnarray}

\section{ Potential Model for 
$Q$ and $\bar Q$ Interacting  at Large Distances} 
\label{sec:con}

We wish to describe the nonperturbative interaction between $Q$ and
$\bar Q$ in terms of a potential model.  Such an interaction is known
for $Q$ and $\bar Q$ in the color-singlet state.  The interaction
consists of an attractive short-range color-Coulomb interaction and
a long-range linear potential.  For $Q$ and $\bar Q$ in a color-octet
state, the short-range color-Coulomb interaction is repulsive.  From
lattice calculations for gluonic excitation in the presence of a
static quark-antiquark pair, the spectrum appears to rise linearly
with the relative separation between $Q$ and $\bar Q$, indicating an
effective linear potential for color-octet states\cite{Mor98}.  We can
write the $Q$-$\bar Q$ potential as
\begin{eqnarray}
\label{eq:con}
V_c(r)= C_c {\alpha_s \over r} +\sigma_{c} r    ~~~~~~~(c=1,8)
\end{eqnarray}
where $C_1=-4/3$, $C_8=1/6$, $\alpha_s$ is the strong interaction
coupling constant, $\sigma_1 \sim 1$ GeV/fm, but $\sigma_8$ is not
known.

The confining potential given in Eq.\ (\ref{eq:con}) is a good
description if the spontaneous production of light quark pairs is not
considered.  However, when a quark and an antiquark pulls far apart,
the string joining $Q$ and $\bar Q$ breaks.  Evidence of string
breaking comes experimentally in the production of a large number of
pions in $e^+ e^-$ annihilation at high energies.  This phenomenon can
be explained as follows. The annihilation of $e^+$ with $ e^-$
produces a $q\bar q$ pair.  As the $q$ and the $\bar q$ pull apart,
the string between the quark and the antiquark breaks up into sections
of strings which manifest themselves as pions \cite{Sjo86,Won94}.
Theoretically, the pair production will occur spontaneously through
the Schwinger mechanism when $Q$ is separated from $\bar Q$ at a
distance $r$ such that $\sigma r$ is greater than the mass of a $q
\bar q$ pair \cite{Sch52}.  When the string breaks, the interaction
between $Q$ and $\bar Q$ will be screened and will become a weak
color-van-del-Waal-type interaction, instead of the strong linear
interaction.  The asymptotic behavior of the $Q$-$\bar Q$ potential is
therefore drastically changed when the spontaneous production of light
quark pairs is allowed.  Instead of the confining behavior as usually
assumed, it is more appropriate to describe the motion of the $Q$ and
$\bar Q$ effectively as having an asymptotically free motion at
$r\rightarrow \infty$, on account of the breakup of the string due to
the spontaneous production of light quark pairs, leading to the
production of a $D$ and a $\bar D$.

The influence of the dynamical light quarks on the $Q$-$\bar Q$ potential
has been investigated theoretically by Laermann and collaborators
\cite{Lae86,Bor89} using lattice gauge theory.  It is found that
including the dynamical quarks in the lattice gauge theory leads to a
potential which deviates from the linear potential.  The modified
interaction can be described in the form of a screened color-Yukawa
potential.  Because of this screening due to dynamical quarks, a more
realistic $Q$-$\bar Q$ potential becomes\cite{Lae86,Bor89}
\begin{eqnarray}
\label{eq:pot}
V_c(r)=  C_c {\alpha_s e^{-\mu r}\over r} +\sigma_{c} r {1-e^{-\mu r}
\over \mu_c r} - {\sigma_c \over \mu_c} ,
\end{eqnarray}
where we have used the reference that the potential vanishes at
$r\rightarrow \infty$.  

Under the potential $V_c(r)$ given above, nonperturbative $Q\bar Q$
states $\psi_{JLS}^c(r)$ can be obtained by solving the Schr\" odinger
equation
\begin{eqnarray}
\left \{ -{1 \over M_Q} \nabla^2   + V_c(r) \right \} 
\psi_{JLS}^c (\bbox{r})
= \epsilon \psi_{JLS}^c (\bbox{r})  ,
\end{eqnarray}
where $\epsilon=m_{JLS}-2M_D$.

In our approximate treatment, we consider a single heavy object which
can travel at all distances, with the provision that in going from
small distances to large distances, the nature of this object changes
from being a charm quark at small distances to becoming an open charm
$D$-meson at large distances.  Thus, the mass of this heavy object
should also change from that of the charm quark mass to a $D$-meson
mass.  We describe such a variation of the mass by the effective mass
\begin{eqnarray}
\label{eq:effm}
M_Q(r)=M_D-{ M_D-M_c \over 1 +\exp\{(r-r_m)/a_m\} },
\end{eqnarray}
where we shall use $r_m=0.8$ fm and $a_m=0.2$ fm to describe the
change in nature of this heavy object at a distance of $r_m \sim
0.8$ fm.

We use such the potential (\ref{eq:pot}) and the effective mass
(\ref{eq:effm}) to study the charmonium bound states and the
resonances in the continuum.  The latter are characterized by an
asymptotic phase shift of $(2n+1)\pi/2$ where $n$ is an integer.  A
reasonable description of the bound states and resonances can be
obtained (Table I) by using the parameters of $\alpha_s=0.3$,
$\sigma_1=1.7$ GeV/fm, $\mu=0.28$ GeV, and $M_c$=1.6 GeV. A potential
similar to (\ref{eq:pot}) has been used successfully to study the
energy spectrum of charmonium states \cite{Din95}.  As in this earlier
work, we find that a large value of the parameter $\sigma_1$ is needed
when the linear interaction is modified to take the form of the
screened potential of Eq.\ (\ref{eq:pot}).  We shall use this
potential to generate the $J/\psi$ and $\psi'$ wave functions for
subsequent studies.

\vskip 1.0cm
\centerline{Table I.  Charmonium states obtained with the screened
potential Eq.\ (\ref{eq:pot})}
\vskip 0.4cm

\begin{tabular}{|c|c|c|c|c|c|} \hline
{\it Bound State}& {\it Resonance}	& {\it Calculated }& {\it Experimental}	&
 {\it Calculated} 	& {\it Experimental} \\
      &    	& {\it Energy} (GeV)& {\it Energy} (GeV)	&
 $\Gamma_{ee^+}$ (keV) 	& $\Gamma_{ee^+}$ (keV) \\
\hline
1S	&	& 3.145			& 3.07			& 8.11
				&  5.26$\pm$0.37 \\
2S	&	& 3.560		        & 3.66			& 2.57
				&  2.14$\pm$0.21 \\
	&  3S	& 3.947			& 4.04			&
				&               \\
1P	&	& 3.476			& 3.52			& 
				&               \\
        &  2P	& 3.971			& 			& 
				&                \\
	&  1D	& 3.801 		& 3.77			& 
				&                \\
\hline
\end{tabular}
\vskip 0.6cm

The behavior of the color-octet $(Q\bar Q)_8$ states can be studied by
looking at the states in the potential (\ref{eq:pot}).  The
color-octet potential is repulsive at short distances.  For
sufficiently large values of $\sigma_8$, there is a potential pocket
at intermediate distances.  Depending on the mass of the quark and the
string tension $\sigma_8$, the pocket of the color-octet potential may
hold a bound state.  The greater the rest mass of the quark, the
greater the chance of a color-octet bound state.  In this respect,
there may be a greater chance of finding a color-octet bound state in
a $b\bar b$ system compared to a $c\bar c$ system.

If the linear potential parameter $\sigma_8$ for color-octet states is
greater than 0.31 GeV/fm, then the potential pocket will hold a bound
$[c\bar c]_8$ state. (The state is bound in the sense that its mass is
less than $2 M_D$, but it is unstable against the emission of a
gluon.)  There is, however, little experimental information on the
color-octet states of either a $c \bar c$ or a $b\bar b$ system.
Compared to color-octet states in the continuum, a bound color-octet
state is distinguished by its relatively longer lifetime, as the only
decay mode is the emission of soft gluons, and the emission rate is
lower for the bound state because the energy difference between the
initial and the final state is less for a bound color-octet state.  A
long-lived color-octet quarkonium will travel a large distance
before its color is neutralized.

It is of interest to examine the signature of a long-lived bound color
octet state. Its only decay channel is the emission of a gluon to come
down to color-singlet $Q\bar Q$ bound states. The emitted gluon will
be converted to a light $q\bar q$ pair.  If the initial color-octet
$c\bar c$ pair has a high kinetic energy relative to its complementary
color-octet partner (which is either a $[q\bar q ]_8$ or a $[q
(qq)]_8$), then the $q\bar q$ from the emitted gluon will form two
chains of light mesons (mostly pions) with the color-octet partner as
shown in Fig.\ 1.

\vspace*{0.4cm}
\epsfxsize=300pt
\includegraphics{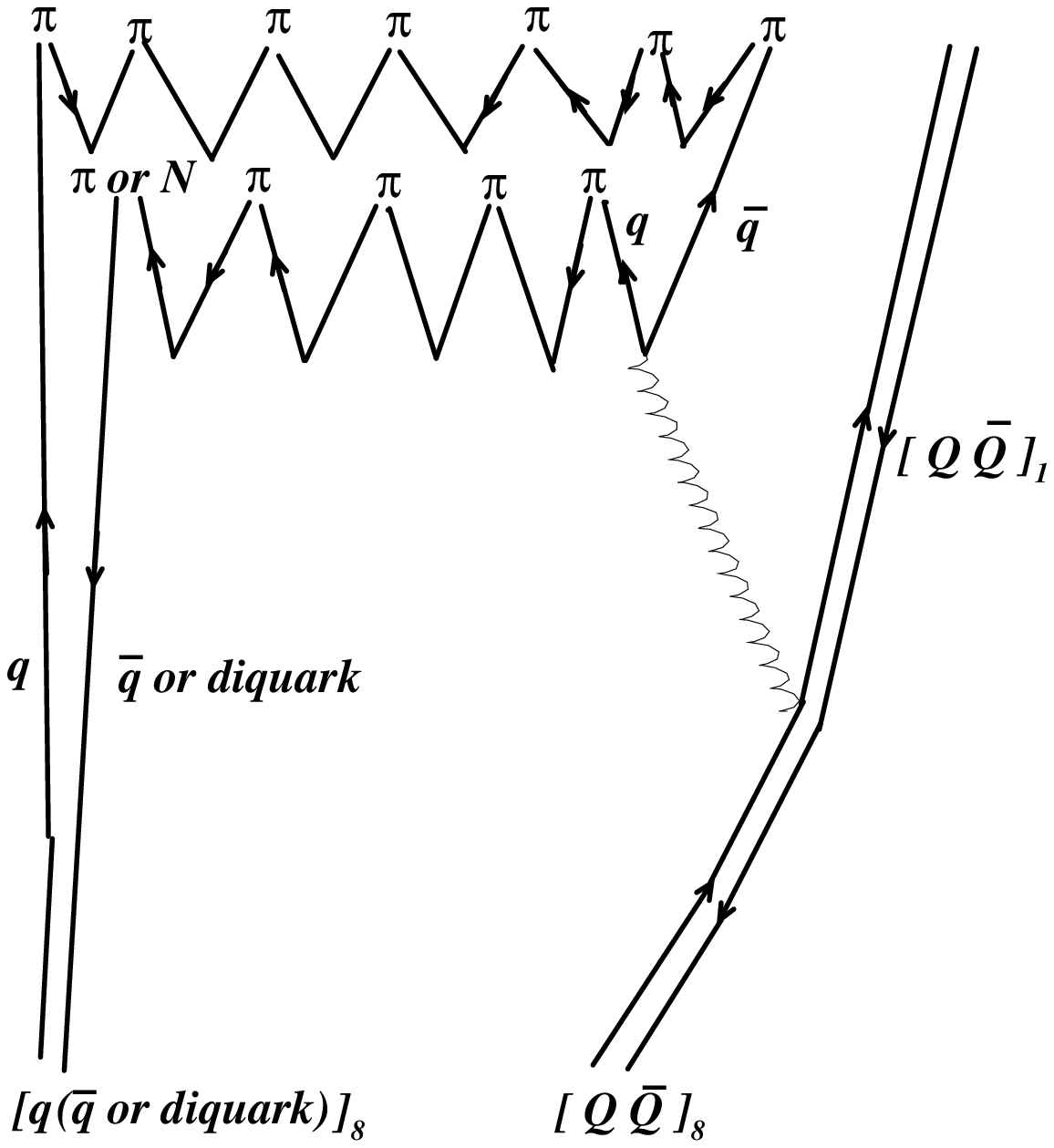}
\vspace*{8.2cm}
\begin{minipage}[t]{14cm}
\noindent {\bf Fig.\ 1}.  {Schematic picture to show how a bound
color-octet $[Q \bar Q]_8$ and its color-octet partner $[q\bar q]_8$
(or $[q (qq)]_8$) evolve into color-singlet objects.}
\end{minipage}
\vskip 4truemm
\noindent 
\vspace*{0.2cm}

In the more general case when the color-octet is accompanied by two
color-octet partners (one from the projectile hadron and one from the
target hadron), the emitted gluon from the $[Q \bar Q]_8$ color-octet
needs to branch out into two gluons, which will be converted into two
pairs of light $[q \bar q]_8$ pairs to form four chains of produced
light mesons.

To search for long-lived color-octet states, it is useful to look at
the kinematic regions where the produced heavy quarkonium is
kinematically separated from the other debris of the collision, as in
the case of $J/\psi$ or $\Upsilon$ at large values of $x_F$.  The
signature of a long-lived color-octet quarkonium state may show up as
a quarkonium at one end accompanied by hadrons which arise
from the breaking of strings joining this color-octet and its
complementary color object at the other end.  There will be chains of
pions in the rapidity gap between the produced quarkonium state and
its complementary partner.  Thus, the search and the measurement of
the chains of hadrons with a leading quarkonium state may reveal the
existence and the characteristics of the quarkonium as a color-octet
state.  In the situation that the pocket is not deep enough to hold a
bound color-octet state, then the intermediate color-octet state
$\psi_{J'L'S'}^{(8)}$, through which the soft gluon is emitted, can
only be in the continuum.

\section{ Continuum States } 

For an intermediate color-octet state in the continuum, the cross
section for the emission of soft gluons depends on its wave function
in two ways.  First, it is needed to give the overlap between the
amplitude from perturbative QCD as given by Eq.\ (\ref{eq:der1}).  It
is also needed to obtain the transition matrix element for soft gluon
emission (given in the next section).

To obtain the continuum wave function, we follow the phase-angle
method discussed in detail by Calogero \cite{Cal67}.
We write the wave function as
\begin{eqnarray}
\psi_{JLS}(\bbox{r}) =R_{JLS}(r)  {\cal Y}_{JLS}(\hat {\bbox{r}})
={u_{JLS}(r) \over kr} {\cal Y}_{JLS}(\hat {\bbox{r}})
\end{eqnarray}
and represent the wave function in terms of the amplitude
$\alpha_{L}(r)$ and the phase shift $\delta_{L}(r)$
\begin{eqnarray}
\label{eq:wf}
u_{JLS}(r)={\alpha_{L}(r) \over \alpha_{L}(\infty)} {\hat D}_{L}(kr) \sin ({{\hat
\delta}_{L} (kr) + \delta_{L}(r)}) \sqrt {{4 \pi \over 2l+1}},
\end{eqnarray}
with the boundary condition that $\delta_{L}(r\rightarrow 0)=0.$ The
functions ${\hat D}(kr)$ and ${\hat \delta}(kr)$ are known
functions\cite{Cal67}:

\begin{eqnarray}
{\hat D}_0(x)=1, ~~~{\hat D}_1(x)=(1+1/x^2)^{1/2},~~
\end{eqnarray}
and
\begin{eqnarray}
{\hat \delta}_0(x)=x; ~~~{\hat \delta}_1(x)=x-\tan^{-1}x.
\end{eqnarray}
The equation for $\delta_{L}(r)$ is\cite{Cal67}
\begin{eqnarray}
\label{eq:edelta}
{d \over dr} \delta_{L}(r)=
- {U(r) \over k} {\hat D}_{L}^2(kr)
\left \{ \sin[{\hat \delta}_{L}(kr)+\delta_{L}(r)]\right \}^2,
\end{eqnarray}
where $U(r)=M_Q(r)V_c(r)+(M_D-M_Q(r))\epsilon.$ After the function
$\delta_{L}(r)$ is evaluated, the amplitude can be obtained from
$\delta_{L}(r)$ by
\begin{eqnarray}
\alpha_{L}(r)=\exp \left \{{1 \over 2k}\int_0^r ds\, U(s)\,
{\hat D}_{L}^2(ks) \, \sin 2[{\hat
\delta}_{L}(ks)+\delta_{L}(s)] \right  \}.
\end{eqnarray}
The numerical integration of Eq.\ (\ref{eq:edelta}) gives the
asymptotic phase shift $\delta(\infty)$ and the continuum wave
function.  The energies at which the asymptotic phase shifts
$\delta(\infty)$ are $(2\times$integer$+1)\pi/2$ are the locations of
the resonances as shown in Table I.

For the wave function and its derivatives at the origin, it is useful
to express the effect of the potential in terms of the $K$-factor, the
analogue of the Gamow factor for the Coulomb interaction \cite{Won97}.
The wave function [Eq.\ (\ref{eq:wf})] near the vicinity of the origin
is given by
\begin{eqnarray}
R_{JLS}(r)={u_{JLS}(r) \over kr}&=&{\alpha_{L}(r) \over \alpha_{L}(\infty)}
j_{L}(kr) \sqrt {{4 \pi \over 2L+1}}     \nonumber\\
& \sim &{\alpha_{L}(r) \over \alpha_{L}(\infty)} 
{ (kr)^{L} \over (2L+1)!!}\sqrt {{4 \pi \over 2L+1}}.
\end{eqnarray}
Therefore, for the factor in Eq.\ (\ref{eq:xsecfn}), we have
\begin{eqnarray}
\sqrt{ {(2L+1)!!\over 4 \pi   \,  {L}!} }  
{ 1 \over  M_Q^L}  { d^{L} R_{JLS}(r) \over dr^L}
 \Biggr |_{r\rightarrow 0}
={\alpha_{L}(r) \over \alpha_{L}(\infty)} 
{k^{L}\over M_Q^{L}}  
\sqrt{ { L! \over (2L+1)!! (2L+1)}}.
\end{eqnarray}
Compared to the case of no interaction for which $\alpha(r)=1$, the
probability is modified by a factor, the $K$-factor,
\begin{eqnarray}
K_L=\left | { \alpha_{L} (r\rightarrow 0 )\over \alpha_{L}(\infty)}
\right |^2.
\end{eqnarray}
In terms of the $K$-factor, the square of the $L$-th derivative of the
wave function at the origin is
\begin{eqnarray}
\label{eq:Lth}
\left |\sqrt{ {(2L+1)!!\over 4 \pi   \,  {L}!} }  
{ 1 \over  M_Q^L}  { d^{L} R_{JLS}(r) \over dr^L}
 \Biggr |_{r\rightarrow 0} \right |^2
=K_L \left |{k\over M_Q} \right |^{2L} 
{ { L! \over (2L+1)!! (2L+1)}}.
\end{eqnarray}
An example of the $K$-factor is the Gamow factor arising from the
Coulomb interaction.  Our numerical $K$-factors can be checked by
noting that for the Coulomb interaction with $V(r)=\alpha/r$, the
$K$-factor can be obtained analytically and shown to be
\begin{eqnarray}
K_L(\eta)= {(L^2+\eta^2)[(L-1)^2+\eta^2] ... (1+\eta^2) \over [L!
(2L+1)!!]^2} {2 \pi \eta \over 1 - \exp\{-2\pi \eta\}}
\end{eqnarray}
where $\eta=\alpha/v$.

\section{ Multipole Transitions } 

We can evaluate the square of the matrix element $|V_{fi}|^2$ for the
emission of soft gluons based on the analogous multipole expansion for
the emission of electromagnetic waves \cite{Ber82,Bla52}.  

\epsfxsize=300pt
\includegraphics{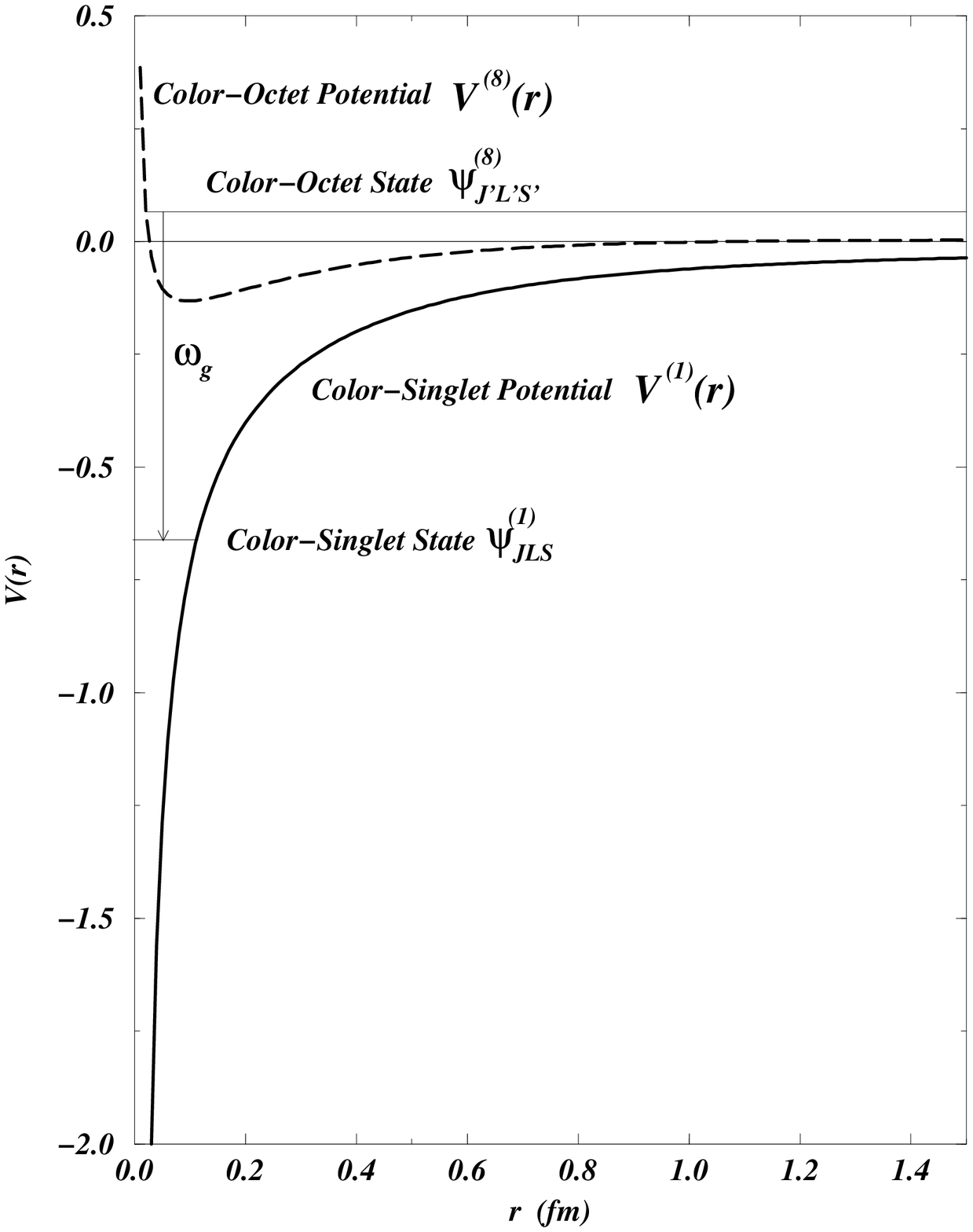}
\vspace*{10.cm}
\begin{minipage}[t]{14cm}
\noindent {\bf Fig.\ 2}.  {Schematic picture of the color-singlet
potential $V^{(1)}(r)$, color-octet potential $V^{(8)}(r)$, the
initial color-octet state $\psi_{J'L'S'}^{(8)}$, and the final
color-singlet state $\psi_{JLS}^{(1)}$ in a soft-gluon emission
process.  }
\end{minipage}
\vskip 4truemm
\noindent 
\vspace*{0.2cm}

We consider the production of a $Q\bar Q$ pair by parton collisions at
an energy $\epsilon$ above the $D\bar D$ threshold.  The $Q\bar Q$ can
be in many different angular momentum states.  After a nonperturbative
time scale, the $Q\bar Q$ pair will evolve into nonperturbative states
$\psi_{J'L'S'}^{(8)}$.  We consider the continuum color-octet states
$\psi_{J'=0,L'=0,S'=0}^{(8)}$ $(^1 S_0^{(8)})$ and
$\psi_{J',L'=1,S'=1}^{(8)}$ ($^3 P_J^{(8)}$) which can decay down to
the color-singlet $\psi_{J=1,L=0,S=1}^{(1)}$ (${}^3S_1^{(1)}$) state
by the emission of a gluon of energy $\omega_g=\epsilon+B$ where $B$
is the binding energy of ${}^3S_1^{(1)}$. (For brevity of notation,
the color state superscript in the wave function $\psi_{JLS}^c$ is
often understood.)  The square of the matrix elements for the emission
of a gluon of energy $\omega_g=\epsilon+B$ from the continuum
color-octet ${Q\bar Q}(^1S_0^{(8)})$ state to the bound color-singlet
$H$=$({}^3S_1^{(1)},S_z=0)$ state through an M1 transition is given by
\cite{Ber82,Bla52}
\begin{eqnarray}
\label{eq:dPM1}
\left |V^{M1}(^1 S_0^{(8)} \rightarrow {}^3S_1^{(1)})  \right |^2
= \biggl ( {8  \over 9} \biggr ) \omega_g^3 | M_{10}' |^2.
\end{eqnarray}

The magnetic dipole transition matrix element $M_{10}'$ is
\begin{eqnarray}
M_{10}'&&={g \over 2M_Q} \sqrt{{ 3 \over 4\pi}} 
\sum_{i=Q, \bar Q} \int d^3 r \psi_{Q\bar Q}^\dagger ({}^1S_0^{(8)}) 
\, \mu_i \sigma_{zi} \,  \psi_{H}({}^3S_1^{(1)}),
\end{eqnarray}
where $\mu_Q$ is the color-magnetic dipole moment of $Q$ and
$\mu_{\bar Q}=-\mu_{Q}\sim 2$.  The $\psi_{Q\bar Q}$ wave function for
the color-octet $Q\bar Q (J'=0,L'=0,S'=0)$ pair in the continuum can be
written as
\begin{eqnarray}
\psi_{Q\bar Q}(J'L'S')
={ u({^1S_0}^{(8)},r)\over kr} {\cal Y}_{J'L'S'}.
\end{eqnarray}
The wave function for the bound
$\psi_{JLS}$ state can be written as
\begin{eqnarray}
\label{eq:Jpsi}
\psi_{JLS} ={ u({}^3 S_1^{(1)}, r) \over r}{\cal Y}_{JLS}.
\end{eqnarray}
From Eqs.\ ({\ref{eq:dPM1})-(\ref{eq:Jpsi}), the matrix element
$M_{10}'$ is given by
\begin{eqnarray}
M_{10}'
={2 g \over M_Q
k} \sqrt{{3 \over 4 \pi}}\int u({^1S_0}^{(8)},r) u({}^3 S_1^{(1)}, r) dr.
\end{eqnarray}
We shall evaluate the above matrix element with wave functions
obtained from the potential model.  As an illustration of the order of
magnitude, we can consider the simple case of
\begin{eqnarray}
u({^1S_0}^{(8)},r)=\sin (kr)\sqrt{4 \pi}
\end{eqnarray}
and the lowest bound $^3S_1$ state wave function as 
\begin{eqnarray}
{u({^3S_1}^{(1)},r) \over r}= 2 \sqrt{\kappa^3} e^{-\kappa r} 
\end{eqnarray}
where $\kappa=\sqrt{M_Q B}$.  Then the square of the matrix element
for M1 gluon radiation from ${}^1S_0^{(8)}$ to ${}^3S_1^{(1)}$ is
\begin{eqnarray}
\label{eq:M1}
\left | V^{M1}(^1 S_0^{(8)} \rightarrow {}^3S_1^{(1)})  \right |^2
={16 g^2 (\mu_Q - \mu_{\bar Q})^2 \omega_g^3  \over M_Q^2}   ~
 { \kappa^5  \over (k^2 +\kappa^2)^4 }.
\end{eqnarray}
The M1 transition rate is constant as $k\rightarrow 0$ because it
involves the spatial overlap of two $S$-state wave functions.

The square of the matrix element for the emission of a gluon of energy
$\omega_g$ from the continuum $(Q\bar Q (^3 P_J^{(8)}),J_z=S_z)$ state
to the bound $H$=$({}^3S_1^{(1)},S_z)$ state through an $E1(m=0)$
transition is \cite{Bla52}
\begin{eqnarray}
\label{eq:dPE1}
\left | V^{E1}(^3 P_J^{(8)} \rightarrow {}^3S_1^{(1)})  \right |^2
=\biggl ({ 8 \over 9} \biggr ) \omega_g^3 | Q_{10} |^2.
\end{eqnarray}
The electric dipole transition matrix element is
\begin{eqnarray}
Q_{10}&=&g \sqrt{3 \over 4\pi} \sum_{i=Q ,\bar
Q} \int d^3 r \Psi_{Q\bar Q}^\dagger (^3 P_J^{(8)}) ~ z_i ~
\Psi_{H}({}^3S_1^{(1)}) 
\nonumber \\
&=& {ig \over k} \sqrt{3 \over 4\pi}   
\int u({^3 P_J^{(8)}},r)~ r ~ 
u({{}^3S_1^{(1)}},r)
dr.
\end{eqnarray}
As an illustration, we shall again evaluate this matrix element with 
a wave function for the non-interacting case
where the $u_{^3 P_J^{(8)}}$ is
\begin{eqnarray}
u({^3 P_J^{(8)}},r)
= \left ( {\sin kr \over k r } -
{\cos kr }\right ) \sqrt{4 \pi \over 3}.
\end{eqnarray}
The square of the matrix
element for E1 gluon radiation from $^3 P_J^{(8)}$ to $^3 S_1^{(1)}$
is
\begin{eqnarray}
\label{eq:E1}
\left | V^{E1}(^3 P_J^{(8)} \rightarrow {}^3S_1^{(1)})  \right |^2
={4\times256 g^2 \over 27}  \omega_g^3  { \kappa^5 k^2 \over
(k^2 +\kappa^2)^6 }.
\end{eqnarray}
The E1 transition rate vanishes as $k \rightarrow 0$.  This arises
because the transition matrix element involves the product of the
$S$-state wave function, which peaks at the origin, with the continuum
$P$-state wave function, which vanishes at the origin.  The product is
zero in the long wavelength ($k$$\rightarrow$0) limit and increases
with increasing kinetic energy.

For the simple case when there is no interaction in the color-octet
state, the ratio of the squares of the  matrix elements for the
production of the lowest $\psi$ state are 
\begin{eqnarray}
\label{eq:rat}
{\left |V^{M1}(^1 S_0^{(8)} \rightarrow {}^3S_1^{(1)})  \right |^2  
\over 
\left | V^{E1}(^3 P_J^{(8)} \rightarrow {}^3S_1^{(1)})  \right |^2}
= { 9 \mu_Q^2  \over 16} {\omega_g^2 \over \epsilon  M_Q}.
\end{eqnarray}
With $\mu_Q=2$, we see that this M1/E1 ratio is large for soft gluons
which carry energies slightly greater than the binding energy $B$.
The $M1$ transition arises from the spin transition current.  In
contrast, the M1 transition arising from the spatial transition currents
would lead to an M1/E1 ratio varying as $v^2=4\bbox{q}^2/M_Q^2$.  In
the present case, because of the angular momentum and spin selection
rules, there is only a contribution to the M1 transition from the spin
transition current, and the usual velocity counting rule based on the
spatial transition currents breaks down \cite{Bla52}.

The ratio of the $|V_{ij}|^2$ for the production of the $\psi'$
state will be more complicated as the $\psi'$ wave function contains
one node and the radial matrix elements will involve cancellation of
contributions of opposite signs.

\section{ Contributions from M1 to E1 Transitions to $J/\psi$ and
$\psi'$ Production} 

From Eq.\ (\ref{eq:xsecfn}), we note that the cross section for going
through different intermediate color-octet states $J'L'S'$ depends
mainly on the product
\begin{eqnarray}
\label{eq:factor}
\left | C_{J'L'S'} \sqrt{ {(2L'+1)!!\over 4 \pi   \,  {L'}!} }  
{ 1 \over  M_Q^{L'}}  \left [ { d^{L'} R_{J'L'S'}(r) \over dr^{L'}}
 \right ]_{r\rightarrow 0} \right |^2 
|V_{J'L'S'\rightarrow JLS}|^2
\left ( {dn \over dm_{J'L'S'}} \right )^2.
\end{eqnarray}
We can consider the production of the $J/\psi (^3S_1)$ state via an M1
transition from the intermediate color-octet state
$\{J'L'S'\}={}^1S_0$.  There can also be an E1 transition from the
intermediate color-octet state $\{J'L'S'\}={}^3P_0$.  To evaluate
various quantities in the above equation, we need a description of the
color-octet state.  There is the short-range repulsive interaction
represented by $C_c=-1/6$ in Eq.\ (\ref{eq:pot}).  If the linear
interaction between the $Q$ and $\bar Q$ is strong enough, a bound
color-octet state may be present in the pocket in the intermediate
distance range to feed to the observed color-singlet state.  The bound
color-octet state will most likely be an $L'=0$ state rather than
$L'=1$ state because the latter lies at a much higher energy.  A bound
color-octet $L'=0$ state will make the transition to the color-singlet
state by an M1 transition, and there will be no $E1$ transition in
this case.

Due to the absence of experimental evidence for bound color-octet
states at present, we shall consider color-octet states to exist only
in the continuum and shall assume only the color-Yukawa interaction
with no linear interaction.  Using such a color-octet potential, we
obtain the wave function in the continuum using methods outlined in
Section \ref{sec:con}.  The $K$-factor can be evaluated to give the
square of the $L'$-th derivative of the wave function of the continuum
color-octet state at the origin [Eq.\ (\ref{eq:Lth})].  The wave
functions in the initial color-octet state and the final color-singlet
state can be used to evaluate the matrix elements of $M1$ and
$E1$ operators.  We then obtain the various quantities in Eq.\
(\ref{eq:factor}).  For the purposes of displaying the results, we
introduce the product $F_R$ defined as
\begin{eqnarray}
F_R= \left | C_{J'L'S'} \sqrt{ {(2L'+1)!!\over 4 \pi   \,  {L'}!} }  
{ 1 \over  M_Q^{L'}}  \left [ { d^{L'} R_{J'L'S'}(r) \over dr^{L'}}
 \right ]_{r\rightarrow 0} \right |^2 .
\end{eqnarray}

There is a limit for the maximum energy of an emitted soft gluon.  One
envisages that gluons with an energy higher than $\omega_g$(cutoff)
will be emitted at a shorter time scale and will be described by
perturbative QCD and not by the soft gluon emission through an
intermediary non-perturbative state.  The cutoff of this soft gluon
energy depends on the size of the bound state, as the time required for
a gluon to travel the length of the radius $R/2$ of a $J/\psi$
corresponds to an energy of $2\hbar/R \sim 0.8$ GeV. (Here, $R$ is the
separation between $Q$ and $\bar Q$ which is about 0.5 fm for
$J/\psi$.)  Therefore the maximum energy of an emitted soft gluon should
be of the order of 0.8 GeV.  The extracted color-octet matrix element
depends on this cutoff gluon energy.  We find that a value of
$\omega_g({\rm max})=0.7$ GeV gives the best description and is
approximately consistent with this estimate of $2\hbar/R$ from the
size consideration.  For this maximum value of soft-gluon energy, the
invariant mass of the color-octet state is 3.8 GeV.

In Fig. 3 we show $|V_{ij}|^2$, $F_R$, and $(dn/dm)^2$ as a function of
the invariant mass $m_{J'L'S'}$ of the intermediate color-octet state
for the production of the $J/\psi$ state.  Corresponding quantities
for the production of $\psi'$ are shown in Fig.\ 4.


\epsfxsize=300pt
\includegraphics{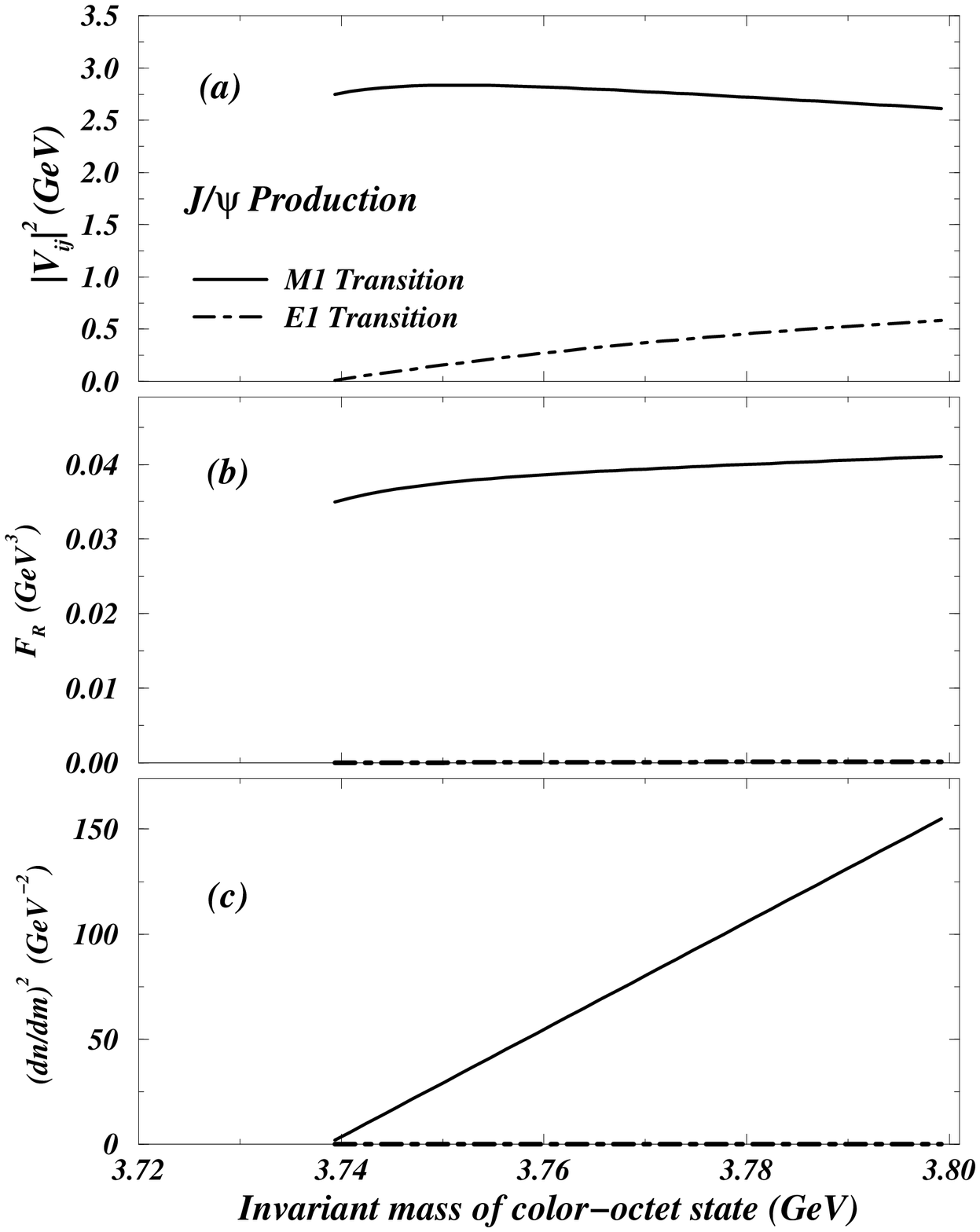}
\null\vspace*{9.7cm}
\begin{minipage}[t]{14cm}
\noindent {\bf Fig.\ 3}.  {$|V_{ij}|^2$, $F_R$, and $(dn/dm)^2$ as a
function of the invariant mass $m_{J'L'S'}$ of the intermediate
color-octet state for the production of $J/\psi$.  }
\end{minipage}
\vskip 4truemm
\noindent 
\vspace*{0.2cm}

For the production of $J/\psi$ with an $M1$ transition, the initial
color-octet state is an $L'=0$ state.  The results of Fig.\ 3 indicate
that the square of the matrix element $|V_{ij}|^2$ and $F_R$ are
approximately constants as a function of the color-octet invariant
mass.  The density of states $dn/dm_{J'L'S'}$ for $L'=0$ increases
with the invariant mass of the color-octet state.  For the $E1$
transition leading to the production of $J/\psi$ , color-octet states
making the $E1$ transition are $L'=1$ states.  The quantity
$|V_{ij}|^2$ increases monotonically from zero as the invariant mass
increases. The factor $F_R$ for E1 transition is small because it is
proportional to $v^2$ (Fig.\ 3$b$).  Because of the centrifugal
barrier, the density of states $dn/dm_{J'L'S'}$ for $L'=1$ is zero for
color-octet states in the range of invariant mass up to 3.8 GeV under
consideration (Fig.\ 3$c$).  A higher energy, $L'(L'+1)/M_QR^2$, is
needed for the $Q$ and $\bar Q$ to overcome the centrifugal barrier to
come to a distance of $R\sim 0.5$ fm for these angular momentum $L'=1$
states.  As a consequence, there is no $E1$ transition from the
color-octet $L'=1$ state to the $J/\psi$ state in the region of soft
gluon emission.  The $M1$ transition dominates over the $E1$
transition for $J/\psi$ production.  The color-octet matrix elements
for various transitions are given in Table II.  We obtain $\langle
{\cal O}_8^{J/\psi}(^3S_1^{(8)}) \rangle = 0.076 {\rm GeV}^3$ and 0
for $\langle {\cal O}_8^{J/\psi}(^3S_1^{(8)}) \rangle$, with the
combination $\langle {\cal O}_8^{J/\psi}(^3S_1^{(8)}) \rangle +
(3/M_Q^2) \langle {\cal O}_8^{J/\psi}(^3S_1^{(8)}) \rangle = 0.076
{\rm GeV}^3$, which can be compared with the value of 0.066 extracted
from the CDF measurements.

\epsfxsize=300pt
\includegraphics{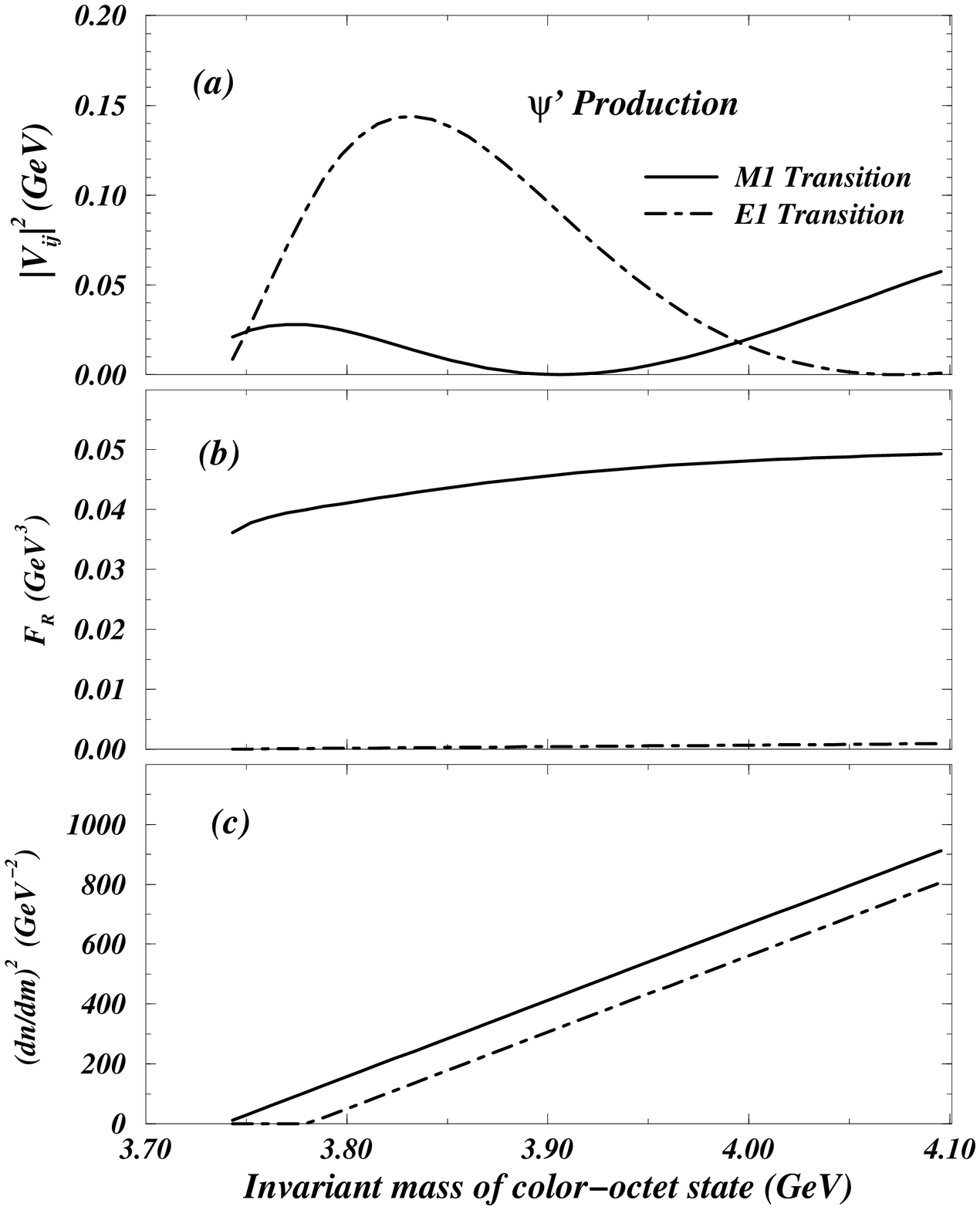}
\null\vspace*{9.4cm}
\begin{minipage}[t]{14cm}
\noindent {\bf Fig.\ 4}.  {$|V_{ij}|^2$ and $F_R$ as a function of the
invariant mass $m_{J'L'S'}$ of the intermediate color-octet state for
the production of the $\psi'$ state.
}
\end{minipage}
\vskip 4truemm
\noindent
\vspace*{0.2cm}

For the production of $\psi'$, the size estimate gives a maximum gluon
energy of the order of $2\hbar/(1 {\rm ~fm})\sim 0.4$ GeV which
corresponds to a maximum invariant mass of about 4.1 GeV for the
color-octet state.  For the production of $\psi'$, the radial wave
function of the final state has a node and a change of the sign of the
wave function.  As a result, the matrix element $V_{ij}$ for the $M1$
transition involves a high degree of cancellation and the magnitude of
the matrix element is small for the $M1$ transition (Fig.\ 4$a$).  For
the M1 transition the $F_R$ factor is approximately a constant but
$(dn/dm_{J'L'S'})^2$ increases monotonically with $m_{J'L'S'}$ (Fig\
4).  For the $E1$ transition $|V_{ij}|^2$ depends on the invariant mass of the
color-octet state and is much larger than that for the $M1$ transition
at $m_{J'L'S'}=3.85$ GeV.  On
the other hand, the $F_R$ factor is proportional to $k^2/M_Q^2$ for
the $E1$ transition and is small in magnitude (Fig\ 4$b$).  The square
of the density of the $L'=1$ states, $(dn/dm)^2$, increases with the
color-octet invariant mass after a threshold.  The combined result is
a small $E1$ transition probability for the production of $\psi'$ as
shown in Table II.  We obtain $\langle {\cal O} _8 ^{\psi'} ({}^1 S_0)
\rangle= 0.015 {\rm ~GeV}^3$ for the $M1$ transition, and $\langle
{\cal O} _8 ^{\psi'} ({}^3 P_0) \rangle/M_Q^2= 0.0001 {\rm GeV}^3$ for
the $E1$ transition.  The $M1$ transition probability is again much
greater than the $E1$ transition probability.  The sum $\langle {\cal
O} _8 ^{\psi'} ({}^1 S_0) \rangle + 3\langle {\cal O} _8 ^{\psi'}
({}^3 P_0) \rangle/M_Q^2 $ agrees approximately with the matrix
element extracted in CDF measurements \cite{Cho96}.

\vskip 0.4cm
\centerline{Table II.  Color-Octet Matrix Elements from Eq.\ (\ref{eq:come})}
\vskip 0.4cm

\begin{tabular}{|c|c|c|c|} \hline
{\it Matrix Element}& {\it Matrix Elements from}& {\it Matrix
Elements}\\
                    & {\it CDF Measurements} & {\it  Calculated with
Eq.\ (\ref{eq:come})} \\
\hline
$\langle {\cal O} _8 ^{J/\psi} ({}^1 S_0) \rangle$
  &               
  & 0.076 ${\rm GeV}^3$    \\
$\langle {\cal O} _8 ^{J/\psi} ({}^3 P_0) \rangle/M_Q^2$
  & 
  & 0                                  \\
\hline
$\langle {\cal O} _8 ^{J/\psi} ({}^1 S_0) \rangle +
3\langle {\cal O} _8 ^{J/\psi} ({}^3 P_0) \rangle/M_Q^2$
  & 0.066 ${\rm GeV}^3$	   
  & 0.076 ${\rm GeV}^3$    \\
\hline
$\langle {\cal O} _8 ^{\psi'} ({}^1 S_0) \rangle$ 
  &
  & 0.015 ${\rm GeV}^3$    \\
$\langle {\cal O} _8 ^{\psi'} ({}^3 P_0) \rangle/M_Q^2 $
  &
  & 0.0001 ${\rm GeV}^3$    \\
\hline
$\langle {\cal O} _8 ^{\psi'} ({}^1 S_0) \rangle + 
3\langle {\cal O} _8 ^{\psi'} ({}^3 P_0) \rangle/M_Q^2 $
  & 0.018 ${\rm GeV}^3$
  & 0.015 ${\rm GeV}^3$    \\
\hline
\end{tabular}
\vskip 0.6cm

It is easy to show that the probability for the transition ${}^1
S_0^{(8)} \rightarrow ({}^3 S_1^{(1)},S_z)$ is the same for
$S_z=-1,0,$ and 1.  Therefore, an initial ${}^1 S_0^{(8)}$ state will
lead to an equal population of the final magnetic substates of ${}^3
S_1^{(1)}$ quarkonium and consequently gives an isotropic angular
distribution of muons from the decay of the quarkonium.

In $J/\psi$ and $\psi'$ production at fixed-target energies, the
fusion of two gluons is the dominant mechanism of direct $J/\psi$ and
$\psi'$ production and gives rise to the color-octet ${}^1 S_0^{(8)}$
and ${}^3 P_J^{(8)}$ states.  Thus, the direct production of $J/\psi$
and $\psi'$ bound states by the emission of very soft gluons occurs
mainly through the M1 ${}^1 S_0^{(8)}\rightarrow {}^3S_1^{(8)}$
transition, and the produced ${}^3 S_1$ bound state will be nearly
unpolarized, with an approximately isotropic angular distribution for
the decay muons.  Assuming $\langle {\cal O}_8^{J/\psi} ({}^3 P_J)
\rangle /M_Q^2=0$ and including the small contributions from the
color-singlet mechanism and $q\bar q$ annihilation, Beneke $et~al.$
\cite{Ben96} obtain $\lambda=0.15$ for the muon angular distribution
$1+\lambda \cos^2 \theta$ for $\psi'$ decay in the Gottfried-Jackson
frame, which falls within the error of the experimental value of
$\lambda=0.02\pm 0.14$ for the $\psi'$ at $\sqrt{s}=21.8$ GeV
\cite{Hei91}.  The experimental $\lambda$ value for direct $J/\psi$
production is not yet available.  The measured $\lambda$ value for
total $J/\psi$ production is 0.028$\pm$0.004 at $\sqrt{s}=15.3$ GeV
\cite{Ake93} and is -0.02$\pm$0.06 for $\sqrt{s}=21.8$ GeV
\cite{Bii87}, which includes direct and indirect $J/\psi$ production
from the decay of $\chi$ states.  The theoretical calculation of
$\lambda$ for the total $J/\psi$ yield is still incomplete because of
the difficulty of the Color-Octet Model to produce the observed
$\chi_1/\chi_2$ ratio \cite{Ben96}.

We now return to the question of the universality of the matrix
elements.  With $\langle {\cal O}_8^{{}^3S_1}({}^1S_0)\rangle >>$
$\langle {\cal O}_8^{{}^3S_1}({}^3 P_J)\rangle/m_c^2$, the discrepancy
between the matrix elements in the CDF measurement and the
fixed-target measurements is reduced to a factor of 2 for $J/\psi$ and
4 for $\psi'$ \cite{Ben96}.  The discrepancy can be further reduced
when one takes into account the physical masses and allows for the
finite energy carried by the soft gluon, as suggested by Beneke
$et~al.$ \cite{Ben96}.

\section{ Conclusions and Discussions } 

For the Color-Octet Model to be a valid description, we need to
understand the soft gluon emission process.  As the soft gluon
emission occurs at a late stage of the production process, the final
state can be described in terms of an interacting $Q$-$\bar Q$
potential. We use a potential model to describe the initial and final
states in the soft gluon emission process.

In our discussions of the long-distance behavior of the $Q$-$\bar Q$
potential, it is necessary to take into account the effects of the
spontaneous production of light quarks as the heavy $Q$ and $\bar Q$
pull apart.  Accordingly, the potential at large distances is screened
due to the breaking of the string. A potential which has such a
property was used successfully to describe the color-singlet
charmonium bound states and resonances.

Not much is known about the potential between $Q$ and $\bar Q$ in a
color-octet state except that it is repulsive at short distances.  The
combined potential may have a pocket deep enough to hold a bound
state.  A bound color-octet state is a relatively long-lived object
having interesting experimental signatures such as the occurrence of 
chains of pions in the rapidity gap between the produced quarkonium
state and its complementary partner as the octet is pulling apart from
its complementary color partner.

In the absence of evidence for the occurrence of color-octet bound
states at present, we consider color-octet states in the continuum.
For our studies, we assume a weak color-octet interaction which
contains only the repulsive color-Yukawa interaction at short
distances to give illustrative results concerning the production
process.

We found that the production cross section depends on three factors:
the magnitude of the $M1$ and $E1$ matrix elements, the wave function
or its derivatives at the origin, and the density of the color-octet
states in the soft-gluon emitting region.  It turns out that for
$J/\psi$ production, the magnitude of the $M1$ matrix elements is
relatively constant for the range of soft-gluon emissions, while the
magnitude of $E1$ transition matrix elements increases monotonically
from zero as the energy of the soft gluon increases.  The E1
transition is, however, highly inhibited because within the range of low
energy soft-gluons (up to $\omega_g \sim 0.8$ GeV), the density of the
initial color-octet $L'=1$ states is zero as a higher energy is needed
to overcome the centrifugal barrier.  As a consequence, the $M1$
transition dominates over the $E1$ transition.

For $\psi'$ production, the transition matrix element for the $M1$
transition is small because of the cancellation involving a final
state whose wave function changes sign in different regions of $r$.
However, the density of $L'=1$ initial color-octet states is small and
limits the contribution of the $E1$ transition. The $M1$ transition
is again the dominant mode of transition.

It is clear from the above discussions that the dominance of the M1
over the E1 transition arises mainly from the limiting factor of the
density of $P$-wave color-octet states.  Thus, the dominance of M1
soft gluon emission over the E1 transition is a rather general
property which is not affected by the strength of the $Q$-$\bar Q$
potential.  On the other hand, the magnitude of the M1 cross section
is sensitive to the maximum soft gluon energy and the $Q$-$\bar Q$
interaction.

In conclusion, while there are questions concerning the Color-Octet
Model, some of these may be resolved by careful refinements of the
details of the model.  In particular, the dominance of the M1
transition ${}^1 S_0^{(8)} \rightarrow {}^3 S_1^{(1)} $ over the E1
transition ${}^3 P_J^{(8)}\rightarrow {}^3 S_1^{(1)} $ can explain the
isotropic muon angular distribution and reduce the discrepancies of
the matrix elements between the CDF and the fixed-target
measurements. There may be the universality of the color-octet matrix
elements when one takes into account the physical masses and the
finite energies of the soft gluon.

\vspace*{-0.5cm}
\section*{Acknowledgments}
\vspace {-0.5cm}

Tha author would like to thank Drs. G. T. Bodwin, M. Beneke, and
H. Crater for helpful discussions.  This research was supported by the
Division of Nuclear Physics, U.S.D.O.E.  under Contract
No. DE-AC05-96OR21400 managed by Lockheed Martin Energy Research Corp.

\end{document}